\documentclass{article}

\usepackage[text={173 mm,25cm},centering,letterpaper]{geometry}
\usepackage{amsmath, amsthm, amssymb}
\usepackage{graphicx}
\usepackage[margin = 10pt, font=small, labelfont=bf, labelsep = endash]{caption}
\usepackage[pdfpagemode=UseOutlines, hidelinks]{hyperref}
\usepackage{apacite}

\begin{document}

\title{Projections of the Transient State-Dependency of Climate Feedbacks}

\author{Robbin Bastiaansen\thanks{Institute for Marine and Atmospheric research Utrecht, Department of Physics, Utrecht University, Utrecht, The Netherlands (r.bastiaansen@uu.nl, h.a.dijkstra@uu.nl, a.s.vonderheydt@uu.nl)} , Henk A. Dijkstra\footnotemark[1] \thanks{Centre for Complex System Studies, Department of Physics, Utrecht University, Utrecht, The Netherlands} , Anna S. von der Heydt\footnotemark[1] \footnotemark[2] }

\maketitle

\begin{abstract}
When the climate system is forced, e.g. by emission of greenhouse gases, it responds on multiple time scales. As temperatures rise, feedback processes might intensify or weaken. Current methods to analyze feedback strength, however, do not take such state dependency into account; they only consider changes in (global mean) temperature and assume all feedbacks are linearly related to that. This makes (transient) changes in feedback strengths almost intangible and generally leads to underestimation of future warming. Here, we present a multivariate (and spatially explicit) framework that facilitates dissection of climate feedbacks over time scales. Using this framework, information on the composition of projected (transient) future climates and feedback strengths can be obtained. Moreover, it can be used to make projections for many emission scenarios through linear response theory. The new framework is illustrated using the Community Earth System Model version 2 (CESM2).
\end{abstract}

\section{Introduction}

Were the climate system free of feedback processes, it would be easy to predict and control the future climate \cite{Arrhenius1896}. Unfortunately, the climate system is peppered with feedback loops~\cite{heinze2019esd, anna2020quantification}. Because of these, climate change can get suppressed or enhanced, making future projections hard~\cite{Sherwood2020}. Therefore, detailed knowledge of all relevant feedback processes is required to accurately assess potential future climates. However, knowledge of the current strengths of climate feedbacks is not enough. Over time, as the climate state changes, the strengths of climate feedbacks also change~\cite{gregory2016variation, marvel2018internal, armour2013time}; for instance, the albedo-increasing effect of ice is only relevant when there (still) is ice.

As the Earth warms, the strength of positive feedbacks increases~\cite{bony2006well}. Therefore, projections based only on current knowledge of climate feedbacks underestimate future climate change -- especially the committed warming that is to come even if zero emission is achieved~\cite{marvel2018internal,goodwin2018time, senior2000time}. To properly assess different emission scenarios, it is crucial to identify all relevant feedback mechanisms and, additionally, to quantify how their strengths change over time.

For future temperature projections with global climate models, currently the focus lies with the following feedback processes~\cite{zelinka2020causes, klocke2013assessment, cess1975global}. First, the Planck radiation feedback suppresses warming due to increased outgoing radiation. Second, the lapse rate feedback also suppresses warming due to an increase in long-wave radiation escaping to space~\cite{sinha1995relative}. Third is the ice-albedo feedback that enhances warming through a less reflective Earth surface~\cite{curry1995sea}. Fourth is the water vapour feedback which boosts warming because of increased atmospheric water vapour content~\cite{manabe1967thermal}. Finally, there is the cloud feedback, describing changes in cloud type and distribution, which could enhance or suppress warming~\cite{cess1987exploratory}.

The effect of a feedback process is quantified by its contribution to the climate system's radiative response to a certain experienced radiative forcing -- in the current Anthropocene primarily caused by greenhouse gas emissions. Specifically, the top-of-atmosphere radiative imbalance $\Delta N$ (that equals zero when the climate system is in equilibrium) is given by the sum of the radiative forcing $F$ and the radiative response $\Delta R$, i.e.
\begin{equation}
	\Delta N(t) = F(t) + \Delta R(t).
\end{equation}
This response $\Delta R$ is the sum of the \emph{feedback contributions} of all relevant feedbacks. So, writing $\Delta R_j$ for the feedback contribution of feedback $j$,
\begin{equation}
	\Delta R(t) = \sum_{j \in \mathcal{F}} \Delta R_j(t),
	\label{eq:radiative_response_summed}
\end{equation}
where the sum is over all feedback processes. Classically, the \emph{feedback strength} $\lambda_j$ (also called \emph{feedback parameter}) is given by the feedback contribution per unit warming, i.e.
\begin{equation}
	\lambda_j := \frac{\Delta R_j(t)}{\Delta T(t)}.
	\label{eq:definition_feedback_strength}
\end{equation}
From this, the top-of-atmosphere radiative imbalance $\Delta N$ can be related to the warming $\Delta T$ as
\begin{equation}
	\Delta N(t) = F(t) + \left( \sum_{j \in \mathcal{F}} \lambda_j \right) \Delta T(t),
\end{equation}
which can be used for climate projections~\cite{gregory2004new}.

However, the climate system responds on many, vastly different time scales~\cite{anna2020quantification}. In particular, not all feedback processes react similarly on all of these time scales. Consequently, there is no linear relationship between all feedback contributions $\Delta R_j$ and the global warming $\Delta T$ \cite{Andrews2015, armour2017energy, knutti2015feedbacks}. Hence, computations of feedback strengths $\lambda_j$ often find these to change over time~\cite{senior2000time, boer2005climate}; they are certainly not constant, as is often implicitly assumed \cite{zelinka2020causes, klocke2013assessment, meraner2013robust}.

We propose here to extend the classical framework to include dynamics on multiple time scales. To this end, feedback strengths should not be computed for the total time series, as in equation~\eqref{eq:definition_feedback_strength}, but they should be separated in (Koopman) modal contributions instead. This makes the time-dependency of the various feedback processes explicit and opens the possibility of projections -- both transient and equilibrium -- for individual feedback contributions under any forcing scenario.

The rest of this paper is structured as follows. In section \ref{sec:MM}, we describe the theoretical framework, define new feedback strength metrics, and explain the computations of the feedback contributions. In sections \ref{sec:results_abrupt_temporal}-\ref{sec:results_abrupt_spatial}, the framework is applied in a study of climate feedbacks in a $1,000$ year long abrupt4xCO2 experiment in the `Community Earth System Model version 2' (CESM2). In section \ref{sec:results_gradual}, projections of feedback contributions are made for and compared against output of CESM2's 1pctCO2 experiment. Finally, section \ref{sec:discussion} contains a summary and discussion.

\section{Method and Model}\label{sec:MM}

\subsection{Method: the new climate feedback framework}

From linear response theory~\cite{proistosescu2017slow, ruelle2009review, aengenheyster2018point}, the change (response) in an observable $O$ due to a certain forcing $g$ is given by
 
\begin{equation}
	\Delta O(t) = \left( G^{[O]} * g \right)(t) = \int_0^t G^{[O]}(t-s)\ g(s) ds,
\end{equation}
 
where $G^{[O]}$ is the Green function for observable $O$. The Green functions in the climate system are well-approximated by a sum of decaying exponentials (the Koopman modes), with only the contribution of the different modes depending on the observable $O$. That is,
 
\begin{equation}
	G^{[O]}(t) = \sum_{m=1}^M \beta_m^{[O]}\ e^{-t / \tau_m},
\end{equation}
 
where the sum is over all $M$ modes, $\tau_m$ is the $m$-th mode's time scale and $\beta_m^{[O]}$ is the contribution of mode $m$ to the observable's Green function. Because of this, following e.g.~\citeA{proistosescu2017slow}, the response of observable $O$ can also be written as
 
\begin{equation}
	\Delta O(t) = \sum_{m=1}^M \beta_m^{[O]}\ \mathcal{M}_m^g(t),
	\label{eq:general_observable_evolution}
\end{equation}
 
where $\mathcal{M}_m^{g}(t) := \int_0^t e^{-(t-s)/\tau_m} g(s) ds$ contains all time-dependencies and $\beta_m^{[O]}$ all observable-dependencies. Since all feedback contributions $\Delta R_j$ and the global warming $\Delta T$ adhere to this formulation, it is possible to define the feedback strength of feedback $j$ per mode $m$ as
 
\begin{equation}
	\lambda_j^m := \frac{ \beta_m^{[R_j]} }{ \beta_m^{[T]} }.
	\label{eq:feedback_strength_per_mode}
\end{equation}
 
Additionally, the instantaneous feedback strength at time $t$ can be defined as the incremental change in feedback contribution $\Delta R_j$ per incremental change in temperature $\Delta T$ (i.e. the local slope of the graph $(\Delta T, \Delta R_j)$ at time $t$):
 
\begin{equation}
	\lambda_j^\mathrm{inst}(t) := \frac{  \frac{d}{dt} \Delta R_j (t)}{ \frac{d}{dt} \Delta T (t)}
	\label{eq:instantaneous_feedback_strength}
\end{equation}
 
Using this framework, it is also possible to estimate properties of the eventual equilibrium state -- if such state exists. For instance, if $g(t) \equiv \bar{g}$, $\mathcal{M}_m^g(t) = \bar{g} \tau_m \left[1 - e^{-t / \tau_m}\right]$, the equilibrium $\Delta O^*$ is given by
 
\begin{equation}
	\Delta O^* = \sum_{m=1}^M \beta_m^{[O]}\ \bar{g}\ \tau_m.
\end{equation}
 
The equilibrium feedback strength $\lambda_j^*$ for feedback $j$ is then defined as
 
\begin{equation}
	\lambda_j^* := \frac{\Delta R_j^*}{\Delta T^*} = \frac{\sum_{m=1}^M \tau_m\ \beta_m^{[R_j]} }{\sum_{m=1}^M \tau_m\ \beta_m^{[T]} }.
\end{equation}
 
A more elaborate treatment of the mathematics of this framework is given in SI-Text S1. Below, we use this framework on two idealised experiments in CESM2 to understand the temporal evolution of (the strength of) climate feedbacks in these scenarios.

\subsection{Model: CESM2 abrupt4xCO2 and 1pctCO2 experiments}\label{sec:method}

We focus on forcing experiments conducted with the model CESM2~\cite{danabasoglu2020community}, specifically on the model's abrupt4xCO2 ($1,000 y$ long) and 1pctCO2 ($150 y$ long) experiments in CMIP6~\cite{eyring2016overview}. The abrupt4xCO2 experiment is used for fitting of the parameters and computations of feedback strengths per mode; the 1pctCO2 experiment is used to test projections made within the presented framework.

As feedback contributions are not outputted directly by any climate model, these have been determined using additional methodology. For this, we have used the radiative kernel approach~\cite{soden2008quantifying, shell2008using}, in which feedback contributions are computed (per grid point at location ${\bf x}$) as
 
\begin{equation}
	\Delta R_j({\bf x}, t) = \frac{\partial \Delta N}{\partial y_j}({\bf x})\ \Delta y_j({\bf x}, t),
\end{equation}
 
where $y_j$ is the relevant (derived) state variable for the $j$-th feedback and $\partial N/\partial y_j({\bf x})$ is the pre-computed radiative kernel. We have used the publicly available CESM-CAM5 kernels from \citeA{pendergrass2018surface, CAM5kerneldata, CAM5kernelsoftware}. There are others available, but differences between kernels are typically small. The procedure for the cloud feedback contribution is somewhat different because of its more nonlinear nature, but is nevertheless computed following common practice~\cite{soden2008quantifying, shell2008using, klocke2013assessment} -- see SI-Text S4 for more details.

\section{Results}

\subsection{Temporal evolution, abrupt4xCO2}\label{sec:results_abrupt_temporal}

From CESM2 model output, feedback contributions $\Delta R_j({\bf x},t)$, top-of-atmosphere radiative imbalance $\Delta N({\bf x},t)$ and near-surface temperature increase $\Delta T({\bf x},t)$ were computed; changes in observables are determined with respect to the temporal average of a control experiment with pre-industrial CO$_2$ levels. From these, globally averaged datasets $\left\langle \Delta R_j \right\rangle(t)$, $\left\langle \Delta N \right\rangle(t)$ and $\left\langle \Delta T \right\rangle(t)$ were derived. These global time series have been fitted to equation~\eqref{eq:general_observable_evolution}, with exception of $\left\langle \Delta N\right\rangle(t)$ that also includes the radiative forcing and is fitted to
 
\begin{equation}
	\left\langle \Delta N \right\rangle(t) = \left\langle F_\textrm{abr} \right\rangle + \left\langle \Delta R \right\rangle(t) =  \left\langle F_\textrm{abr} \right\rangle + \sum_{m=1}^M \beta_m^{[R]} \mathcal{M}_m^{g_\textrm{abr}}(t),
\end{equation}
 
where the subscript `abr' indicates that the forcings are from the abrupt4xCO2 experiment. More details of the procedure can be found in SI-Text S2.

\begin{figure}
	\centering
	\includegraphics[width=\textwidth]{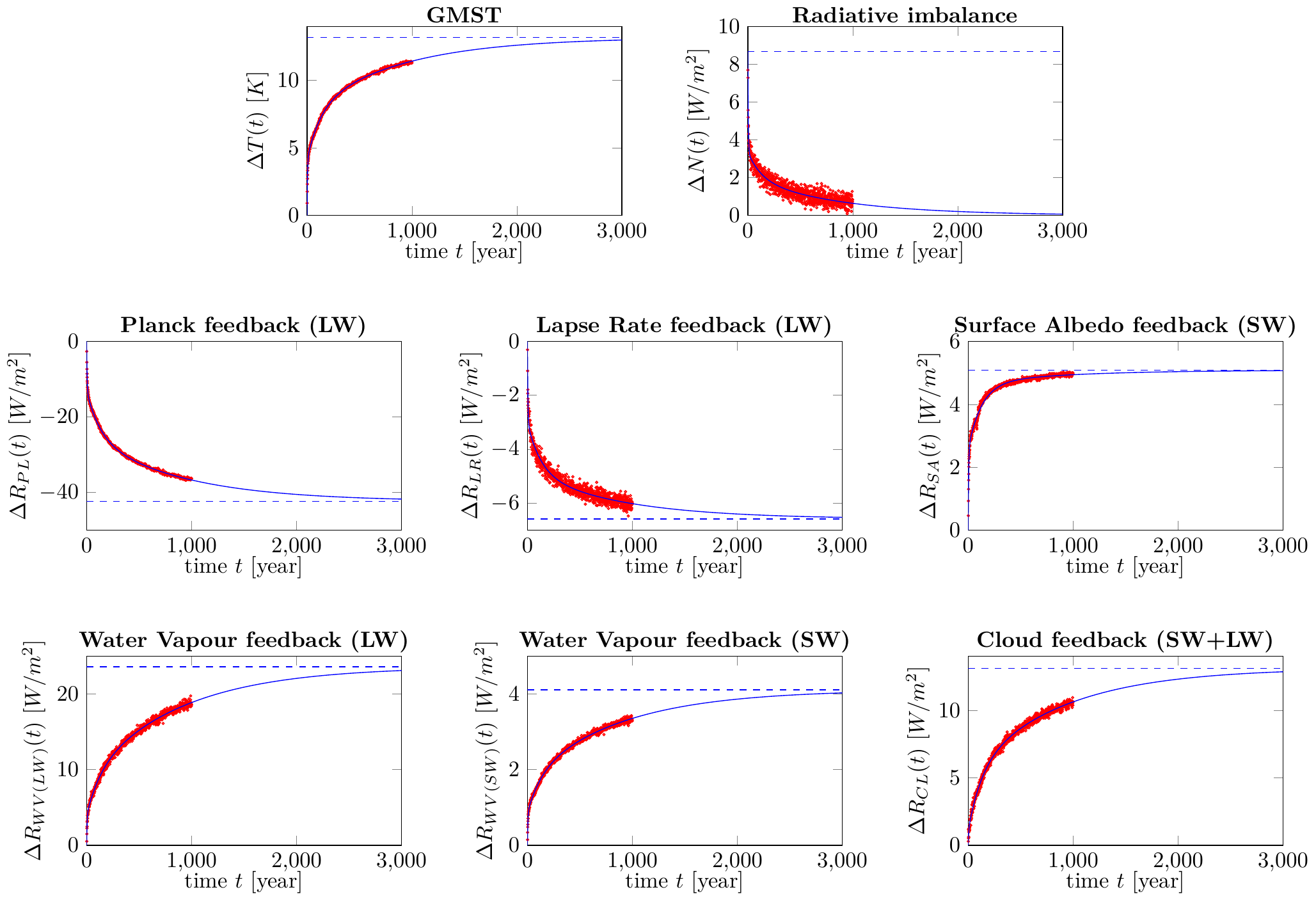}
\caption{Evolution of temperature increase $\Delta T$, radiative imbalance $\Delta N$ and feedback contributions $\Delta R_j$ for the CESM2 abrupt4xCO2 experiment. Red circles denote data points; blue lines the obtained fits. The dashed lines indicate the estimated equilibrium values (or initial forcing in case of the radiative imbalance plot). SI-Table S1 contains the values for the fitted parameters.}
\label{fig:CESM2_fit_data}
\end{figure}

First, the amount of modes $M$ needs to be determined. Tests with various values for $M$ have indicated $M = 3$ is the best choice here: $M < 3$ does not capture all dynamics and $M > 3$ has no significant benefits; see SI-Figure S1. This is also in agreement with other reports that typically find two or three relevant modes~\cite{caldeira2013projections, tsutsui2017quantification, proistosescu2017slow, bastiaansen2020multivariate}. With $M = 3$ fixed, the other parameters are fitted with nonlinear regression. The values for the fitted parameters are given in SI-Table S1. In Figure~\ref{fig:CESM2_fit_data}, the datasets are shown along with the fits and projections, showing a very good resemblance.

\begin{figure}
	\centering
	\includegraphics[width=\textwidth]{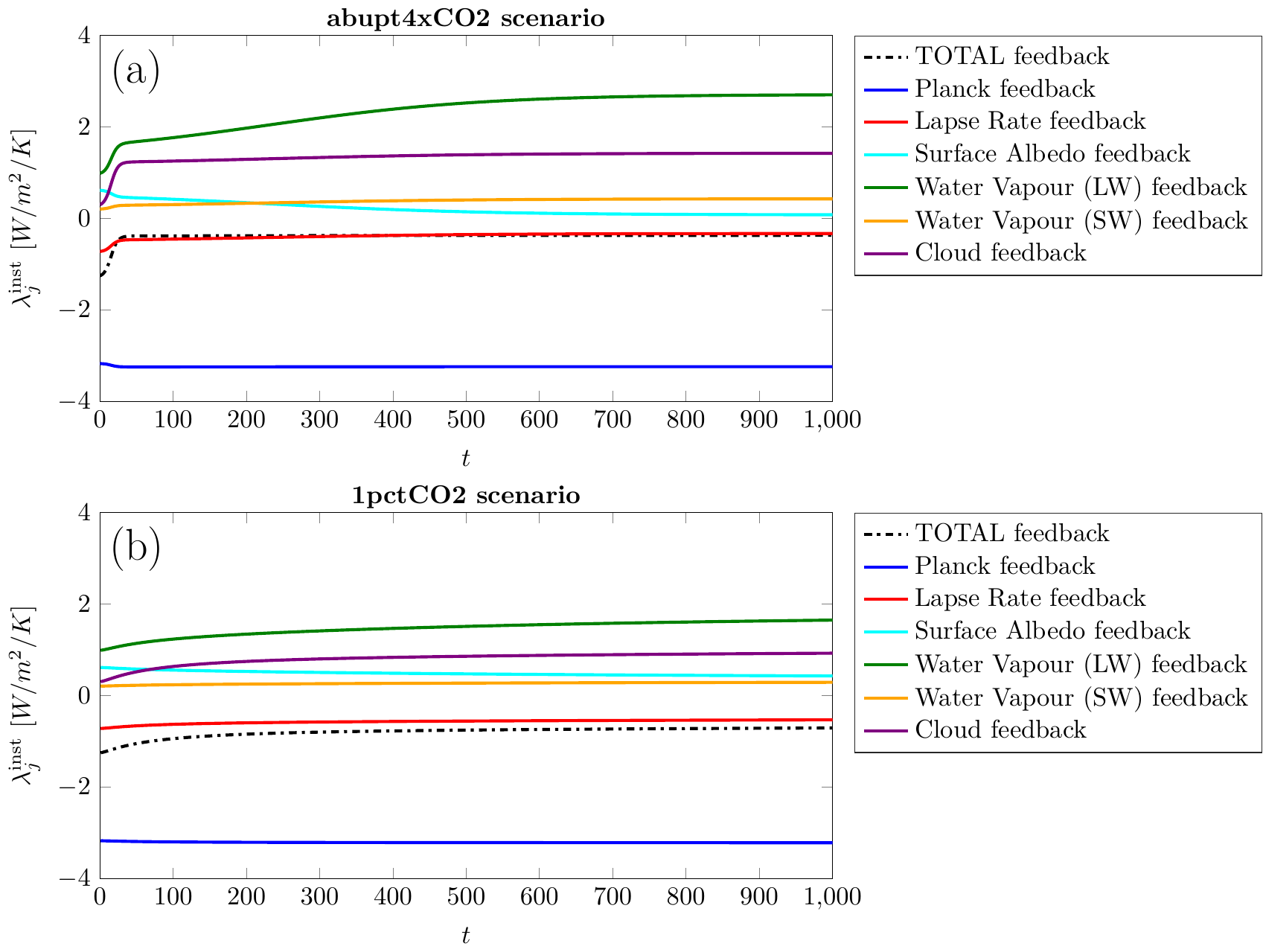}
	\caption{The evolution of climate feedback strengths over time in (a) the abrupt4xCO2 experiment and (b) the 1pctCO2 experiment. Instantaneous feedback strengthts $\lambda_j^\mathrm{inst}(t)$ are computed via equation~\eqref{eq:instantaneous_feedback_strength} from the fits derived from the abrupt4xCO2 experiment (and show projected feedback strengths under a continued $1\%$ yearly CO2 increase).}
	\label{fig:instantaneous_feedback_strengths}
\end{figure}

From the fitted parameters, the feedback strengths per mode are computed using equation~\eqref{eq:feedback_strength_per_mode}. In Table~\ref{tab:CESM2_feedback_parameters}, the results are shown per mode and for the equilibrium values. A plot of the change in (instantaneous) feedback strength over time is given in Figure~\ref{fig:instantaneous_feedback_strengths}A. This shows behaviour on three distinct time scales: (1) a short time scale $\tau_1 = 4.5 y$, (2) an intermediate time scale $\tau_2 = 127 y$ and (3) a long time scale $\tau_3 = 889 y$. The shorter time scale is in agreement with other studies~\cite{zelinka2020causes}, and long-term behaviour seems very plausible. The Planck feedback does not change much over time. The lapse rate feedback becomes weaker over time since temperatures increase. The surface albedo feedback diminishes as sea-ice is melting. The water vapour feedback, however, strengthens as warmer air can contain more water, which -- as demonstrated here -- does lead to an almost tripling in this feedback's strength over time. Finally, the cloud feedback becomes more dominant over time. The estimated equilibrium values are a bit higher generally compared to other studies (e.g.~\citeA{zelinka2020causes}) since longer time scales are incorporated in the present study, which leads to higher temperature values and more positive total feedback strength~\cite{proistosescu2017slow, rugenstein2020equilibrium, marvel2018internal, goodwin2018time}.

Since the individual feedback contributions $\Delta R_j$ should sum to the total radiative response $\Delta R$, feedback strengths should satisfy $\lambda^m = \sum_{j \in \mathcal{F}} \lambda_j^m$. However, because of ignored feedback interactions, imperfect radiative kernels and other limitations of the feedback contribution computations, a mismatch of up to $10\%$ is deemed inconsequential. Looking at the feedback parameters per mode in Table~\ref{tab:CESM2_feedback_parameters}, we see a good balance for the intermediate time scale ($-0.02\ Wm^{-2}K^{-1}$ imbalance), but not for the short ($+0.54\ Wm^{-2}K^{-1}$ imbalance) or long time scales ($-1.46\ Wm^{-2}K^{-1}$ imbalance). Taking the large uncertainty into account for the shortest time scale, this imbalance might be insignificant still. However, for the long time scale, the imbalance is large and even the signs are not the same. This points to a missing negative feedback on this time scale (or inaccuracies in the computational methods for the feedback contributions $\Delta R_j$). Because the equilibrium values show a smaller imbalance ($-0.43\ Wm^{-2}K^{-1}$), the alleged missing feedback (or feedbacks) seem to be especially important during the transient state.

\begin{table}
\caption{Values for the climate feedback parameters $\lambda_j^m$ per mode. These have been computed from the fitted parameters (SI-Table S1) via equation~\eqref{eq:feedback_strength_per_mode}. Time scales $\tau_m$ have units `year'; feedback parameters $\lambda_j^m$ have units `$W / m^2 / K$'. Plusminus values indicate $95\%$ confidence intervals, which were propagated from the $95\%$ confidence intervals for the fitted parameters assuming a normal distribution for all errors and no existing correlations between parameters.}
\label{tab:CESM2_feedback_parameters}
\centering
\newcommand{\fitreport}[2]{$#1$ {\scriptsize ($\pm\ #2$)}}
\begin{tabular}{r||ccc||c}
					& Mode 1 				& Mode 2				& Mode 3				& Equilibrium 			\\\hline
$\tau_m$ 			& \fitreport{4.5}{0.1}		& \fitreport{127}{3.8}	& \fitreport{889}{50}		& ---						\\\hline
$\lambda_m$ 		& \fitreport{-1.28}{0.08}	& \fitreport{-0.38}{0.03}	& \fitreport{-0.37}{0.02}	& \fitreport{-0.66}{0.03}	\\\hline
Planck (LW)			& \fitreport{-3.16}{0.02}	& \fitreport{-3.24}{0.02}	& \fitreport{-3.23}{0.01}	& \fitreport{-3.21}{0.05}	\\
Lapse Rate (LW)		& \fitreport{-0.73}{0.03}	& \fitreport{-0.50}{0.03}	& \fitreport{-0.32}{0.03}	& \fitreport{-0.50}{0.01}	\\
Surface Albedo (SW)	& \fitreport{+0.62}{0.04}	& \fitreport{+0.56}{0.02}	& \fitreport{+0.08}{0.10}	& \fitreport{+0.39}{0.01}	\\
Water Vapour (LW)	& \fitreport{+0.97}{0.03}	& \fitreport{+1.38}{0.02}	& \fitreport{+2.71}{0.01}	& \fitreport{+1.79}{0.04}	\\
Water Vapour (SW)	& \fitreport{+0.21}{0.09}	& \fitreport{+0.26}{0.05}	& \fitreport{+0.43}{0.02}	& \fitreport{+0.31}{0.01}	\\
Clouds (SW + LW)	& \fitreport{+0.27}{0.36}	& \fitreport{+1.19}{0.02}	& \fitreport{+1.43}{0.01}	& \fitreport{+1.00}{0.03}	\\\hline
sum					& \fitreport{-1.82}{0.37}	& \fitreport{-0.36}{0.07}	& \fitreport{+1.09}{0.11}	& \fitreport{-0.22}{0.08}	\\\hline
residue				& \fitreport{+0.54}{0.38}	& \fitreport{-0.02}{0.08}	& \fitreport{-1.46}{0.11}	& \fitreport{-0.43}{0.08}
\end{tabular}

\end{table}

\subsection{Spatial distribution, abrupt4xCO2}\label{sec:results_abrupt_spatial}

With a slight modification of~\eqref{eq:feedback_strength_per_mode}, the spatial structure of the (Koopman) modes can be included. Following a similar procedure in~\citeA{proistosescu2017slow}, we take
 
\begin{equation}
	\Delta R_j({\bf x},t) = \sum_{m=1}^M\ \beta_m^{[R_j]}({\bf x})\ \mathcal{M}_m^{g_\mathrm{abr}}(t),
\end{equation}
 
and similar formulas for $\Delta N({\bf x},t)$ and $\Delta T({\bf x},t)$. More details can be found in SI-Text S2.2.

The space-dependent $\beta_m^{[R_j]}({\bf x})$ encompasses the spatial structure of the observable $\Delta R_j$ in the $m$-th mode. Again, it is also possible to estimate equilibrium properties, leading to an estimation of the spatial distribution of feedback contributions and warming in equilibrium. Using the time scales $\tau_m$ from the fits of the global observables, the spatial coefficients $\beta_m^{[R_j]}({\bf x})$ can be computed efficiently using linear regression methods.

\begin{figure}
	\centering
	\includegraphics[width=\textwidth]{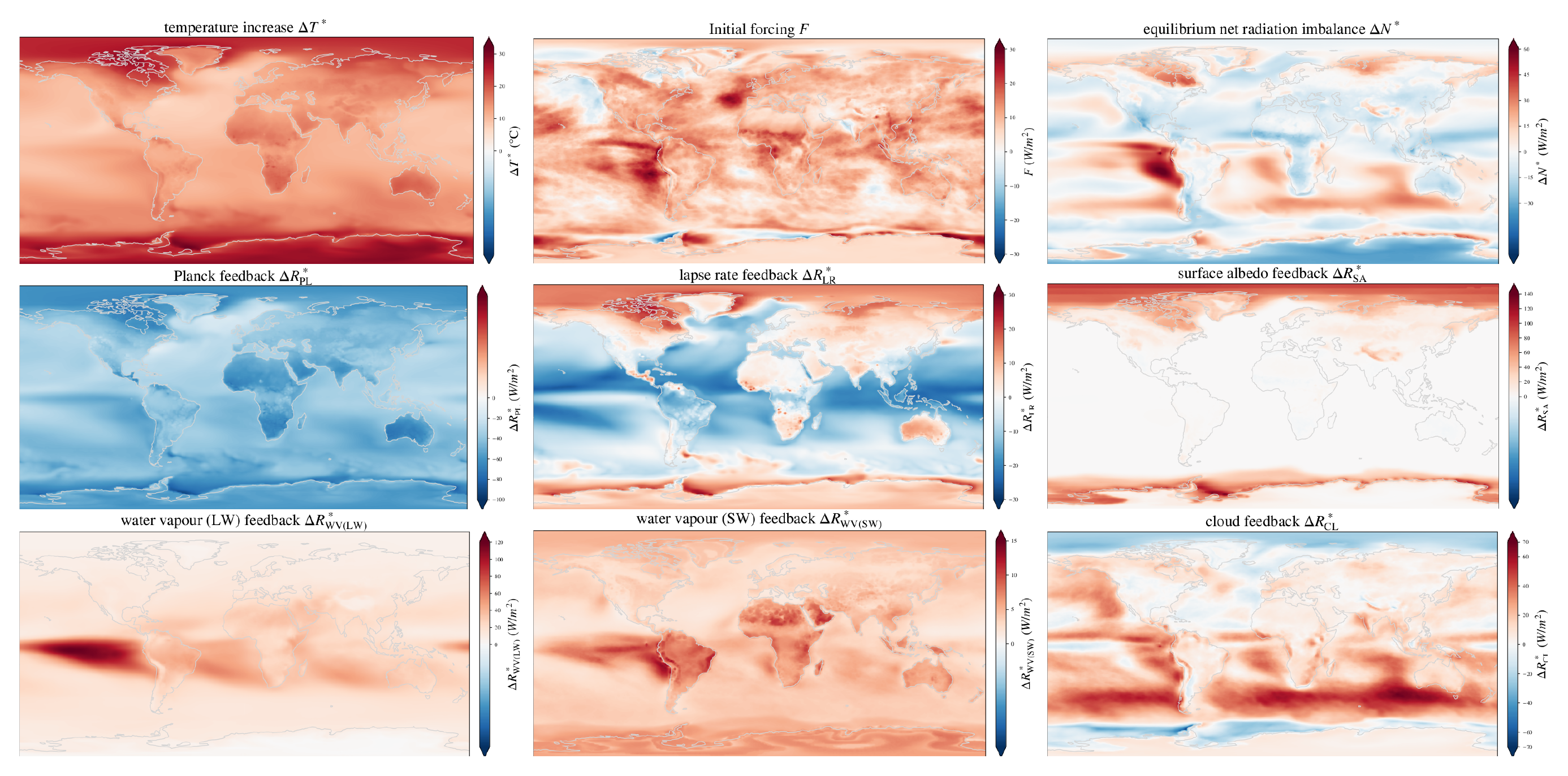}
	\caption{Estimated spatial distribution of warming $\Delta T_*(\bf{x})$, radiative imbalance $\Delta N_* (\bf{x})$ and feedback contributions $\Delta R_j^*(\bf{x})$ in equilibrium, and initial (effective) radiative forcing $F(\bf{x})$ for the CESM2 abrupt4xCO2 experiment. Equilibrium distributions are derived from the fitted spatial modes; initial forcing $F(\bf{x})$ is obtained directly from fits.}
	\label{fig:spatial_equilibrium}
\end{figure}

In SI-Figures S3--S10, the spatial modes of all feedback contributions, warming and radiative imbalance are given. The estimated equilibrium distributions are shown in Figure~\ref{fig:spatial_equilibrium}. The patterns of feedback contributions are in agreement with previous studies \cite{proistosescu2017slow, dong2020intermodel, armour2013time, boeke2020nature, andrews2018dependence, dessler2013observations, soden2008quantifying, Andrews2015}, although they seem more pronounced here -- probably due to a better incorporation of long-term behaviour. Near the poles, temperature increase gets close to $+30^{\circ}C$ (compared to $+12^{\circ}C$ near the equator) and large changes (up to $+135\ Wm^{-2}$) in surface albedo are found related to sea-ice melting. Moreover, the lapse rate feedback contribution increases by about $10\ Wm^{-2}$ in these polar regions. Near the equator, the water vapour feedback gets much stronger (more than $+50 Wm^{-2}$) over the oceans -- especially in the Pacific cold tongue where changes up to $+120 Wm^{-2}$ are found. Cloud feedback changes mostly over southern hemisphere oceans (up to $+69\ Wm^{-2}$), and northern hemisphere changes are smaller (on average only $+2.7\ Wm^{-2}$) compared to previous studies \cite{andrews2018dependence, dong2020intermodel, armour2013time}.

\subsection{Projections for 1pctCO2 experiment}\label{sec:results_gradual}

In the 1pctCO2 experiment, the forcing ($g_\textrm{grad}$) is given by a yearly $1\%$ increase in atmospheric CO$_2$. Replacing $\mathcal{M}_m^{g_\mathrm{abr}}$ by $\mathcal{M}_m^{g_\mathrm{grad}}$ in~\eqref{eq:feedback_strength_per_mode}, while using the values for $\tau_m$ and $\beta_m$ as fitted with the abrupt4xCO2 experiment, gives a projection for feedback contributions and warming in the 1pctCO2 experiment. This procedure, following linear response theory~\cite{ruelle2009review, lucarini2011statistical, ragone2016new}, has been used for warming projections already in many models, including global climate models~\cite{lembo2020beyond, aengenheyster2018point, maier1987transport, hasselmann1993cold}. Although typically used on an ensemble of model runs and for smaller forcings, we use it here nevertheless to illustrate its potency. More details can be found in SI-Text S3.

\begin{figure}
	\centering
	\includegraphics[width=\textwidth]{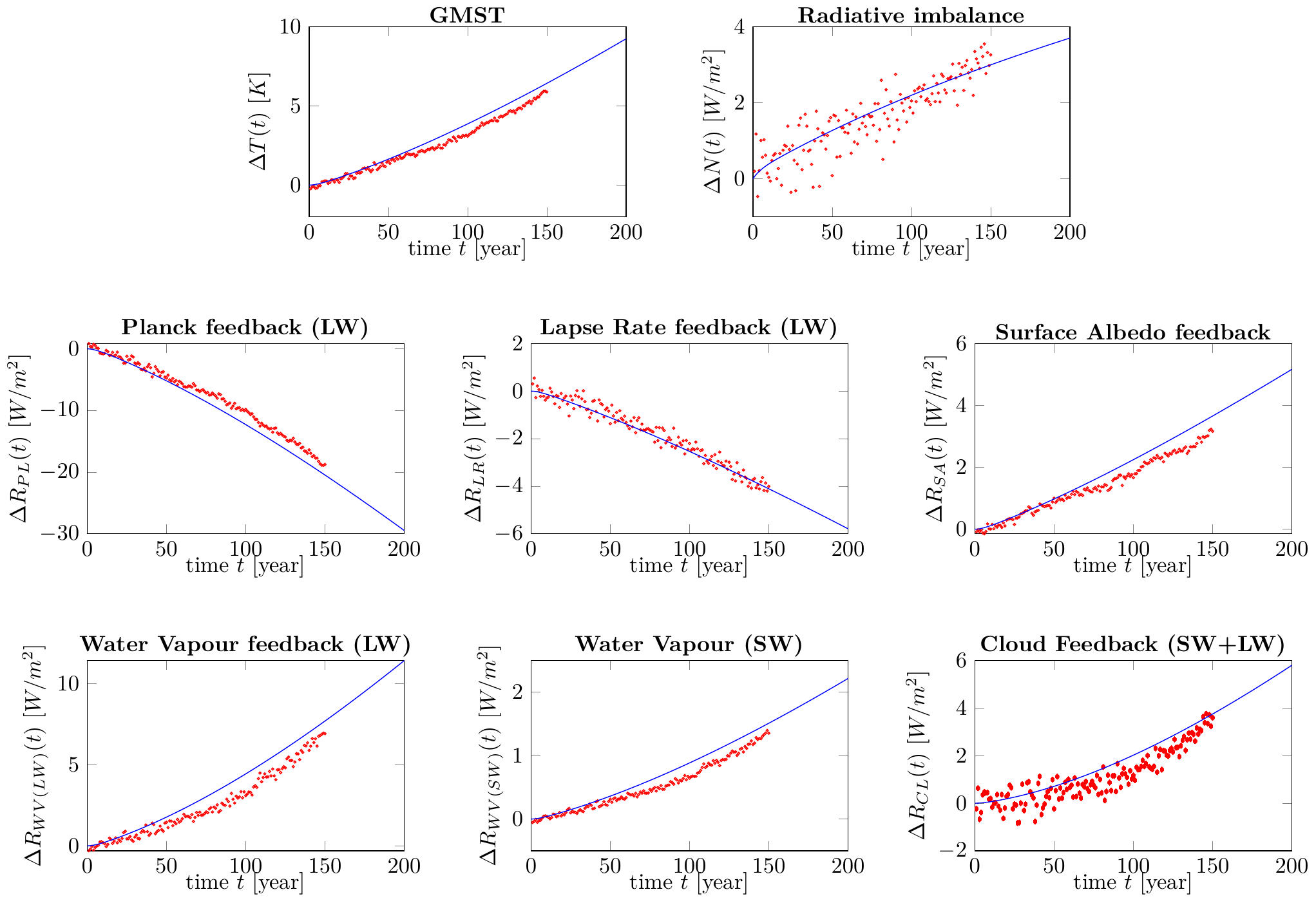}
	\caption{Projections and actual values for globally averaged feedback contributions $\Delta R_j$, warming $\Delta T$ and top-of-atmosphere radiative imblance $\Delta N$ in the CESM2 1pctCO2 experiment. Red circles denote data points obtained from model output. The lines show projections based on data from the abrupt4xCO2 experiment using the introduced framework.}
	\label{fig:1pct_global_projection}
\end{figure}

In Figure~\ref{fig:1pct_global_projection}, projections for global feedback contributions are shown along with data coming directly from the CESM2 1pctCO2 experiment. Qualitatively, the projections are able to reproduce the actual trends in the globally averaged observables. The quantitative values overestimate feedback contributions by $10\%-20\%$ (less for the lapse rate feedback; more for cloud and long-wave water vapour feedbacks up to about $t=120 y$), which seems reasonable for this set-up. Projections for the temporal evolution of instantaneous feedback strengths in this scenario are given in Figure~\ref{fig:instantaneous_feedback_strengths}B. These show lasting changes over centuries -- primarily in (long-wave) water vapour feedback, surface albedo feedback and cloud feedback -- and a less negative total feedback parameter, indicating the climate system becomes more sensitive to radiative forcing over time.

Using the same approach, it is also possible to make projections for the evolution of the spatial distributions. In SI-Figures~S11--S18 the projections for the feedback contributions $\Delta R_j$, the radiative imbalance $\Delta N$ and the temperature increase $\Delta T$ at $t = 140 y$ are shown. Videos showing these results for other times are also available (SI-videos). From all of these, it can be seen that the spatial projections reproduce the general larger spatial patterns and even some of the smaller features, although the internal variability is of course not present in the projections. The spatial patterns are similar to those found in the abrupt4xCO2 experiment (Figure~\ref{fig:spatial_equilibrium}) and show the same kind of response in polar and equatorial regions. In general, the projections seem to overestimate the spatial feedback contributions, similar to the overestimation of the global feedback contributions.

\section{Summary and discussion}\label{sec:discussion}

In this study, we have introduced a new multivariate feedback framework that enables analysis and projections of individual climate feedbacks, including their spatial distribution, over different time scales. This framework deviates from the common practice of linearly regressing feedback contributions to global warming~\cite{zelinka2020causes, marvel2018internal, klocke2013assessment, meraner2013robust} -- which is not valid, especially when considering long-term effects~\cite{Andrews2015, armour2017energy, knutti2015feedbacks}. Instead, these contributions are here taken into account directly, and temporal evolution is explicitly considered via a nonlinear regression to obtain (Koopman) modes for each feedback. These modes, indicating behaviour over different time scales, encapsulate the climate system's dynamics and can thus be used both for analysis and projections of climate feedbacks -- even for other forcing scenarios.

Using the presented framework, we have analyzed an abrupt4xCO2 experiment in CESM2. This clearly showed evolution of climate feedbacks over time, such as a diminishing of surface albedo feedback in the polar regions as sea-ice melts. Strikingly, the water vapour and cloud feedbacks showed a large increase over time: water vapour feedback, primarily over oceans around the equator, almost tripled in strength over a 1,000 year period and cloud feedback mainly changed over southern hemisphere oceans. Analysis also showed that the (commonly) considered feedback processes do not sum up to the total feedback for long time scales, pointing to a missing negative feedback on these time scales. Since estimated equilibrium feedback strengths do balance, the alleged missing feedback (or feedbacks) is expected to play a role primarily during the transient. As such, the ocean heat uptake is a good candidate to fill this gap. However, as of now, no radiative kernels exist for the ocean heat uptake's effect on the top of atmosphere radiative balance, making it impossible to test this hypothesis. Nevertheless, tracking its effect over time seems necessary to accurately understand the long-term transient behaviour.

Another benefit of tracking the temporal evolution of climate feedbacks is that this makes it possible to project the climatic change for all sorts of emission scenarios without the need to have dedicated model experiments for them all. In this paper, we have shown the capabilities of such projections by comparing them to a (single) 1pctCO2 experiment with the same CESM2 model. The projections did capture the qualitative trends, but also showed an overestimation for almost all global and spatial climate feedback contributions. However, this possibly is due to internal variability associated with a single run. Hence, results are expected to be even better quantitatively when a (potentially small) ensemble of runs is used, as application of similar linear response techniques on e.g. temperature response have shown previously~\cite{lembo2020beyond, aengenheyster2018point}.

As climate models become more and more detailed, these models also become more computationally expensive. Therefore, it is practically impossible to compute climate responses for many forcing scenarios and/or on very long time scales. As climate feedbacks become more positive with rising temperatures, it is necessary to track these changes to prevent underestimation of climate change and adequately account for committed (long-term) warming. Hence, both extrapolation techniques and response theory can play a significant role to alleviate the gaps. The multivariate feedback framework in this study can contribute to this, allowing for long-term estimation and climate projections for different forcing scenarios. Moreover, as these projections are multivariate, they can inform us on more than global mean warming alone, and also indicate how a potential future climate state may behave differently compared to our current one.

\section*{Data statement}
Radiative kernels from~\citeA{pendergrass2018surface, CAM5kerneldata, CAM5kernelsoftware} have been used; the kernels were downloaded from \url{https://doi.org/10.5065/D6F47MT6} and accompanying software from \url{https://doi.org/10.5281/zenodo.997899}.

CESM2 data has been downloaded on the fly from Google's cloud storage mirror of CMIP6 data using the `intake-esm' utility package in Python (\url{https://doi.org/10.5281/zenodo.4243421}).

\section*{Acknowledgements}
All MATLAB and Python codes are made available on \url{https://github.com/Bastiaansen/Climate_Feedback_Projections}. This project is TiPES contribution \#114: this project has received funding from the European Union's Horizon 2020 research and innovation programme under grant agreement 820970.

\bibliography{sources}

\end{document}


\maketitle

\tableofcontents

\listoffigures

\listoftables

\newpage
\linenumbers


\section{Tracking the temporal evolution of Climate Feedbacks}

The Earth is not a closed system, because energy can enter and leave the atmosphere in the form of radiation. The net amount of the top-of-atmosphere radiative imbalance is typically denoted by $\Delta N$ and equals the amount of incoming shortwave solar radiation minus the outgoing short- and longwave radiation (i.e. $\Delta N = N_{SW,\downarrow} - N_{SW,\uparrow} - N_{LW,\uparrow}$). The precise amounts of these outgoing radiative fluxes is influenced by any forcing (e.g. due to CO$_2$ emissions) and by the many feedback processes of the Earth system (e.g., sea-ice melting, changes in surface albedo and cloud formation). Hence, the radiative imbalance $\Delta N$ can be seen as a function of any applied forcing $\mu$ and the system state $\vec{y}$; that is,
\begin{linenomath*}\begin{equation}
	\Delta N = \Delta N\left(\vec{y};\mu\right). \label{eq:TOAimbalance_general}
\end{equation}\end{linenomath*}
The Earth system is said to be in (radiative) equilibrium when $\Delta N = 0$. So, for a given (constant) forcing $\mu = \mu^{A}$, an equilibrium state $\vec{y} = \vec{y}^A$ has to satisfy
\begin{linenomath*}\begin{equation}
\Delta N\left(\vec{y}^A; \mu^A\right) = 0.
\label{eq:TOAimbalance_unperturbed}
\end{equation}\end{linenomath*}
If the amount of forcing were to change, say $\mu(t) = \mu^{A} + \Delta \mu(t)$ (not necessarily as a function of time), the system state $\vec{y}(t)$ and the radiative imbalance $N(t)$ will respond and change over time. Denoting the new system state by $\vec{y}(t) = \vec{y}^{A} + \Delta \vec{y}(t)$, a substitution in \eqref{eq:TOAimbalance_general} gives the new imbalance as
\begin{linenomath*}\begin{equation}
	\Delta N(t) = \Delta N\left(\vec{y}^A + \Delta \vec{y}(t); \mu^A + \Delta \mu(t)\right).
\end{equation}\end{linenomath*}
If the perturbation $\Delta \mu(t)$ is small enough, the response $\Delta y(t)$ will also typically be small. In this case, this expression can be approximated via a Taylor expansion as a (multi-)linear function. Specifically,
\begin{linenomath*}\begin{equation}
	\Delta N(t) = \Delta N\left(\vec{y}^A; \mu^A\right) + \frac{\partial \Delta N}{\partial \mu}\left(\vec{y}^A; \mu^A\right) \Delta \mu(t) + \frac{\partial \Delta  N}{\partial \vec{y}}\left(\vec{y}^A; \mu^A\right) \Delta \vec{y}(t) + h.o.t.
	\label{eq:TOAimbalance_linearised}
\end{equation}\end{linenomath*}
In this expression, $\Delta N\left(\vec{y}^A;\mu^A\right)$ is the radiative imbalance in the unperturbed state, which equals zero per \eqref{eq:TOAimbalance_unperturbed}. The term $\frac{\partial \Delta N}{\partial \mu}\left(\vec{y}^A; \mu^A\right) \Delta \mu(t)$ denotes the direct effect of the forcing and is typically called the radiative response $F(t)$. The resulting radiative response $\Delta R(t)$ to this forcing is given by $\frac{\partial \Delta N}{\partial \vec{y}}\left(\vec{y}^A; \mu^A\right) \Delta \vec{y}(t)$, which includes all the climate feedback processes. Finally, $h.o.t.$ stands for `higher order terms' and includes all non-linear and interaction terms that are assumed to be small~\citep{soden2008quantifying, shell2008using}.

The radiative response $\Delta R(t)$ is formed by a combination of all relevant climate feedbacks. The prime ones typically considered are Planck feedback, lapse rate feedback, surface albedo feedback, water vapour feedback and cloud feedback~\citep{zelinka2020causes, soden2008quantifying, shell2008using, klocke2013assessment, meraner2013robust}. For the mathematical framework here, we denote the set of all relevant climate feedbacks by $\mathcal{F}$ (which thus might include other, or undiscovered, feedbacks). Then $\Delta R$ is given by the sum over all these feedbacks as
\begin{linenomath*}\begin{equation}
	\Delta R(t) = \sum_{j \in \mathcal{F}} \Delta R_j(t) := \sum_{j \in \mathcal{F}} \frac{\partial \Delta N}{\partial y_j}\left(\vec{y}^A;\mu^A\right) \Delta y_j(t).
\end{equation}\end{linenomath*}
So, $\Delta R_j(t)$ denotes the contribution over time of the $j$-th feedback to the full radiative response of the (linearised) climate system. Henceforth, we will refer to these as \emph{feedback contributions}.

In climate studies, typically, the feedback contributions are not reported; instead, values for \emph{feedback parameters} (or \emph{feedback strengths}) are given: the feedback contribution per Kelvin warming. Mathematically, these are defined as
\begin{linenomath*}\begin{equation}
	\lambda_j(t) := \frac{\Delta R_j(t)}{\Delta T(t)},
	\label{eq:climate_feedback}
\end{equation}\end{linenomath*}
where $\Delta T(t)$ denotes the change in global mean surface temperature (GMST). Moreover, normally also a full climate feedback parameter $\lambda$ is defined as
\begin{linenomath*}\begin{equation}
	\lambda(t) := \frac{\Delta R(t)}{\Delta T(t)},
\end{equation}\end{linenomath*}
which -- if all feedback processes are taken into account -- leads to $\lambda (t) = \sum_{j \in \mathcal{F}} \lambda_j(t)$ (with possibly a small inconsistency because of the neglected higher order and interaction terms in \eqref{eq:TOAimbalance_linearised}). Here, we have explicitly denoted the time-dependency of $\lambda_j(t)$. However, often, the feedback parameters are interpreted as constants in time. In practice, their values are even inferred from transient simulation data via linear regression of $\Delta R_j(t) = \lambda_j \Delta T(t)$ \citep[e.g.][]{zelinka2020causes, marvel2018internal, klocke2013assessment, meraner2013robust}.

However, such linearity assumption does not hold -- especially when incorporating climate dynamics over longer time scales \citep{Andrews2015, armour2017energy, knutti2015feedbacks}. In general, a linear dynamical system that is perturbed by some change in forcing $g(t)$ (e.g., $g(t) = \Delta \mu(t)$), follows relatively simple dynamics: the change of any (global mean) observable $O$ evolves as a sum over (Koopman) eigenmodes \citep{proistosescu2017slow}, i.e.
\begin{linenomath*}\begin{equation}
	\left\langle \Delta O \right\rangle (t) = \sum_{m=1}^M\ \beta_m^{[O]}\ \mathcal{M}^g_m(t),
	\label{eq:observable_evolution}
\end{equation}\end{linenomath*}
where $M$ is the amount of (relevant) modes, $\beta_m^{[O]}$ the contribution of the $m$-th mode and $\mathcal{M}_m^g(t)$ the $m$-th mode's response to the applied forcing $g$. Concretely -- assuming the system has only real (negative) eigenvalues -- the eigenmodes are of the form $\beta_m e^{-t / \tau_m}$ where $\tau_m$ is the associated (relaxation) time scale and $\mathcal{M}_m^g$ is given by the (truncated) convolution
\begin{linenomath*}\begin{equation}
	\mathcal{M}^g_m(t) := \left(\ (\ s \mapsto e^{-s / \tau_n}\ ) * g\ \right) (t) = \int_0^t e^{-s / \tau_n}\ g(t-s)\ ds.
\end{equation}\end{linenomath*}

Expression \eqref{eq:observable_evolution} is the common solution to any non-autonomous linear ordinary differential equation. Specifically, the Green function for observable $O$ is given by
\begin{linenomath*}\begin{equation}
	G^{[O]}(t) = \sum_{m=1}^M \beta_m^{[O]} e^{-t / \tau_m},
\end{equation}\end{linenomath*}
and \eqref{eq:observable_evolution} could alternatively be written as
\begin{linenomath*}\begin{equation}
	\left\langle \Delta O \right\rangle (t) = \left( G^{[O]} * g \right)(t) = \int_0^t G^{[O]}(s)\ g(t-s)\ ds.
\end{equation}\end{linenomath*}

As $\Delta R_j$ and $\Delta T$ (and $\Delta R$) follow \eqref{eq:observable_evolution}, $\lambda_j$ in \eqref{eq:climate_feedback} cannot be a constant (unless $M = 1$ or in the case the Green functions for $\Delta R_j$ and $\Delta T$ are identical up to a multiplicative factor).

At the same time, \eqref{eq:observable_evolution} invites the study of the system's dynamics per mode. Hence, following \eqref{eq:observable_evolution} we write
\begin{linenomath*}
\begin{align}
	\Delta R_j(t) & = \sum_{m=1}^M R_j^m(t) := \sum_{m=1}^M \beta_m^{[R_j]} \mathcal{M}^g_m(t); \\
	\Delta R(t) & = \sum_{m=1}^M R^m(t) := \sum_{m=1}^M \beta_m^{[R]} \mathcal{M}^g_m(t); \\
	\Delta T(t) & = \sum_{m=1}^M T^m(t) := \sum_{m=1}^M \beta_m^{[T]} \mathcal{M}^g_m(t).
\end{align}
\end{linenomath*}
Here, $R_j^m$ denotes the feedback contribution for mode $m$ (which thus acts on a timescale $\tau_m$), $R^m$ the complete radiative response at the $m$-th mode and $T^m$ the (global mean surface) temperature at this mode.

We propose here to quantify climate feedback strengths per mode. That is, we define the climate feedback parameters $\lambda_j^m$, which denote the strengths of the $j$-th feedback at mode $m$ as
\begin{linenomath*}\begin{equation}
	\lambda_j^m  := \frac{R_j^m (t)}{T^m (t)} = \frac{ \beta_m^{[R_j]} \mathcal{M}^g_m(t)}{ \beta_m^{[T]} \mathcal{M}^g_m(t)} = \frac{ \beta_m^{[R_j]} }{ \beta_m^{[T]} }, \label{eq:definition_lambda_jm}
\end{equation}\end{linenomath*}
and the total feedback parameters $\lambda^m$ for the full radiative response at mode $m$ as
\begin{linenomath*}\begin{equation}
	\lambda^m  := \frac{R^m(t)}{T^m(t)} = \frac{ \beta_m^{[R]} \mathcal{M}^g_m(t) }{ \beta_m^{[T]} \mathcal{M}^g_m(t)} = \frac{ \beta_m^{[R]} }{ \beta_m^{[T]} }.
\end{equation}\end{linenomath*}

Similar to the classical definitions of the feedback parameters, per mode the individual feedback parameters should again sum up to the total feedback parameter, i.e. $\lambda^m = \sum_{j \in \mathcal{F}} \lambda_j^m$ (again small inconsistencies are possible due to ignored higher order and/or interaction terms in \eqref{eq:TOAimbalance_linearised}).

It is also possible to estimate the instantaneous feedback strength of a feedback at time $t$. For this, the local slope of the graph $(\Delta T(t), \Delta R_j(t))$ at time $t$ can be used, which is computed as the fraction of the derivatives, i.e.
\begin{linenomath*}\begin{equation}
	\lambda_j^\mathrm{inst} (t) = \frac{ \Delta R_j'(t) }{ \Delta T'(t)},
	\label{eq:inst_fb}
\end{equation}\end{linenomath*}
where the primes denote derivatives with respect to time.

Finally, if the applied forcing is constant (i.e. $g(t) = \Delta \mu(t) \equiv \Delta \mu$) and eigenmodes are decaying over time, this formalism also allows for the estimations of equilibrium feedback strengths $\lambda_j^*$ and $\lambda^*$ as
\begin{linenomath*}
\begin{align}
	\lambda_j^* & := \frac{ \lim_{t \rightarrow \infty} R_j(t) }{ \lim_{t \rightarrow \infty} \Delta T(t) } = \frac{ \sum_{m=1}^M \beta_m^{[R_j]} }{ \sum_{m=1}^M \beta_m^{[T]} }; \\
	\lambda^* & := \frac{ \lim_{t \rightarrow \infty} R(t) }{ \lim_{t \rightarrow \infty} \Delta T(t) } = \frac{ \sum_{m=1}^M \beta_m^{[R]} }{ \sum_{m=1}^M \beta_m^{[T]} }. \label{eq:definition_lambda_jstar}
\end{align}
\end{linenomath*}
We stress that these equilibrium properties have no straightforward connection to the transient feedback parameters $\lambda_j^m$ and $\lambda^m$.


\newpage

\section{Climate Feedbacks in CESM2 abrupt4xCO2 experiment}

The above described framework has been applied to CESM2's CMIP6 abrupt4xCO2 model run \citep{eyring2016overview}. This run is performed with CESM2 version 2.1.0 (CAM6; CLM5; POP2; MARBL; CICE5.1; CISM2.1) \citep{danabasoglu2020community}. To compute the feedback contributions $\Delta R_j(t)$, radiative kernels made by \cite{pendergrass2018surface} with CESM-CAM5 were employed~\citep{CAM5kerneldata, CAM5kernelsoftware}. Computations followed the standard process \citep{soden2008quantifying, shell2008using}: radiative kernels were multiplied (per grid cell) with time series for the relevant variable field to obtain location-dependent time series $\Delta R_j({\bf x},t)$, from which global mean time series $\langle \Delta R_j \rangle (t)$ were derived. In this way, we have investigated the following climate feedbacks: (1) Planck feedback, (2) lapse rate feedback, (3) surface albedo feedback, (4) water vapour feedback and (5) cloud feedback. The latter (cloud feedback) has not been computed directly with a radiative kernel, but is indirectly inferred from differences between clearsky and fullsky feedback contributions of the other feedbacks, following \cite{pendergrass2018surface, soden2008quantifying}; more details are given in section~\ref{sec:cloudFeedback}. In addition, also (global mean) time series $\langle \Delta T \rangle (t)$ for surface temperature and $\langle \Delta N \rangle(t)$ for net top-of-atsmophere radiative imbalance have been computed.

In this abrupt4xCO2 experiment, the applied forcing is constant (i.e. $g_\mathrm{abr}(t) = \Delta \mu(t) \equiv \Delta \mu$). Therefore, $\mathcal{M}^{g_\mathrm{abr}}_m(t)$ can be computed exactly as
\begin{linenomath*}\begin{equation}
	\mathcal{M}_m^{g_\mathrm{abr}}(t) = \Delta \mu\ \tau_m\ \left( 1 - e^{- t / \tau_m} \right).
	\label{eq:Mmg_for_abrupt}
\end{equation}\end{linenomath*}
Hence, time series for $\langle \Delta T \rangle (t)$, $\langle N \rangle (t)$ and all feedback contributions $\langle R_j \rangle (t)$ should satisfy
\begin{linenomath*}
\begin{align}
	\langle \Delta T \rangle (t)
		 = \sum_{m=1}^M\ \beta_m^{[T]}\ \Delta \mu\ \tau_m\ \left( 1 - e^{-t / \tau_m} \right) &= \sum_{m=1}^M\ \tilde{\beta}_m^{[T]}\ \left( 1 - e^{-t / \tau_m} \right); \label{eq:CESM2_globalTemperature}\\
	\langle \Delta R_j \rangle (t)
		 = \sum_{m=1}^M\ \beta_m^{[R_j]}\ \Delta \mu\ \tau_m\ \left( 1 - e^{-t / \tau_m} \right) &= \sum_{m=1}^M\ \tilde{\beta}_m^{[R_j]}\ \left( 1 - e^{-t / \tau_m} \right); \\
	\langle \Delta  N \rangle (t)
		 = \langle F_\mathrm{abr} \rangle + \sum_{m=1}^M\ \beta_m^{[R]}\ \Delta \mu\ \tau_m\ \left( 1 - e^{-t / \tau_m} \right) &= \langle F_\mathrm{abr} \rangle+ \sum_{m=1}^M\ \tilde{\beta}_m^{[R]}\ \left( 1 - e^{-t / \tau_m} \right), \hspace{0.3cm} \left(\sum_{m=1}^M \tilde{\beta}_m^{[R]} = - \langle F_\mathrm{abr} \rangle \right), \label{eq:CESM2_globalImbalance}
\end{align}
\end{linenomath*}
where $\tilde{\beta}_m^{[O]} := \beta_m^{[O]}\ \Delta \mu\ \tau_m$. Again, there is a small difference for the cloud feedback -- see section~\ref{sec:cloudFeedback}. So, based on all the time series, values for the time scales $\tau_m$ (applying to all observables) and for mode contributions $\tilde{\beta}_m^{[O]}$ (applying only to observable $O$) can be found from a fit -- for a given $M$. In this study, we have used non-linear least-squares regression for this.

\subsection{Global feedback parameters}

\begin{figure}[t]
\centering
	\includegraphics[width = 0.7\textwidth]{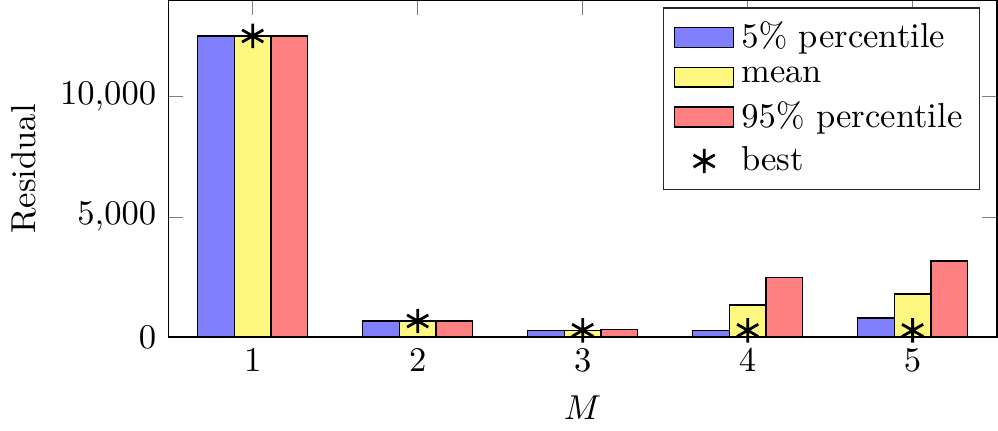}
\caption[Residuals for fits with different amount of modes $M$]{Residuals for fits of \eqref{eq:CESM2_globalTemperature}-\eqref{eq:CESM2_globalImbalance} to global data from CESM2's abrupt4xCO2 experiment run for various values for the amount of modes $M$. Because of the nonlinearity of this regression, found fits can be sensitive to initial guesses. Hence an ensemble of $1,000$ fits (per $M$ value) have been performed with random initial guesses for $\tilde{\beta}_m^{[O]}$ (and $\tau_m$ spaced evenly between $1$ and $1,000$). Blue respectively red bars indicate $5$ respectively $95$ percentile values, yellow bars denote the mean values and stars give the residual for the best fit of the ensemble.}
\label{fig:mode_amount}
\end{figure}

When fitting \eqref{eq:CESM2_globalTemperature}-\eqref{eq:CESM2_globalImbalance} to the (global) data, some value for the number of modes $M$ needs to be specified. Since the true amount of modes is unknown, we have tested values up to $M = 5$ -- see Figure~\ref{fig:mode_amount}. As $M = 3$ seemed to produce consistent fits with low residuals we have chosen $M = 3$ for the fits in this study.

The thus obtained fits are plotted in Main Text Figure 1 along the data. In Table \ref{tab:CESM2_global_parameters}, the values for all fitted parameters are given. From these, the feedback parameters per mode $\lambda_j^m$, as well as equilibrium values $\lambda_j^*$, have been computed; these are reported on in Main Text Table 1.

\begin{table}
\caption[Fitted parameter values for the CESM2 abrupt4xCO2 experiment]{Parameter fits for global data of CESM2's abrupt4xCO2 run in CMIP6. Plusminus values indicate $95\%$ confidence intervals. Estimated values in equilibrium are computed as sums of the modes. $95\%$ confidence intervals have been propagated from the fitted parameters assuming a normal distribution for all errors and no correlations exist between parameters. Not shown in this table is the fitted value for $F_\mathrm{abr}^{cs}$, which was found to be $+8.59$ ({\scriptsize $\pm \ 0.42$}). Note also that $F_\mathrm{abr} = -\sum_{m=1}^3 \tilde{\beta}_m^{[R]}$. Time scales $\tau_m$ carry units `years', the other parameters have units `$ W / m^2$'.}
\label{tab:CESM2_global_parameters}
\centering
\newcommand{\fitreport}[2]{$#1$ {\scriptsize ($\pm\ #2$)}}
\begin{tabular}{r||ccc||c}
						& Mode 1 					& Mode 2 					& Mode 3 					& Equilibrium 				\\\hline\hline
$\tau_m$ 				& \fitreport{4.5}{0.1}			& \fitreport{127}{3.8}		& \fitreport{889}{50}			& 							\\\hline\rowcolor{gray!10}
$\tilde{\beta}_m^{[\textrm{T}]}$			& \fitreport{+04.15}{0.06}	& \fitreport{+03.65}{0.13}	& \fitreport{+05.38}{0.13}	& \fitreport{+13.18}{0.19}	\\\rowcolor{gray!20}
$\tilde{\beta}_m^{[\textrm{R}]}$			& \fitreport{-05.32}{0.32}		& \fitreport{-01.39}{0.11}		& \fitreport{-01.96}{0.08}		& \fitreport{-08.68}{0.34}	\\\rowcolor{gray!10}
$\tilde{\beta}_m^{[\textrm{PL}]}$		& \fitreport{-13.13}{0.10}		& \fitreport{-11.82}{0.28}		& \fitreport{-17.38}{0.22}		& \fitreport{-42.34}{0.37}		\\\rowcolor{gray!20}
$\tilde{\beta}_m^{[\textrm{LR}]}$		& \fitreport{-03.01}{0.06}		& \fitreport{-01.82}{0.11}		& \fitreport{-01.75}{0.11}		& \fitreport{-06.58}{0.17}		\\\rowcolor{gray!10}
$\tilde{\beta}_m^{[\textrm{SA}]}$		& \fitreport{+02.59}{0.06}	& \fitreport{+02.06}{0.10}	& \fitreport{+00.44}{00.12}	& \fitreport{+05.08}{0.17}	\\\rowcolor{gray!20}
$\tilde{\beta}_m^{[\textrm{WV(LW)}]}$	& \fitreport{+04.00}{0.07}	& \fitreport{+05.03}{0.22}	& \fitreport{+14.57}{0.22}	& \fitreport{+23.61}{0.32}	\\\rowcolor{gray!10}
$\tilde{\beta}_m^{[\textrm{WV(SW)}]}$	& \fitreport{+00.86}{0.06}	& \fitreport{+00.93}{0.11}	& \fitreport{+02.32}{0.15}	& \fitreport{+04.11}{0.17}	\\\rowcolor{gray!20}
$\tilde{\beta}_m^{[\textrm{Cloud}]}$	& \fitreport{+01.11}{0.31}	& \fitreport{+04.33}{0.15}	& \fitreport{+07.67}{0.14}	& \fitreport{+13.11}{0.37}	\\
\end{tabular}

\end{table}

The instantaneous feedback strengths $\lambda_j^\mathrm{inst}(t)$ for the abrupt4xCO2 can be computed via~\eqref{eq:inst_fb}. By~\eqref{eq:CESM2_globalTemperature}-\eqref{eq:CESM2_globalImbalance}, this yields
\begin{linenomath*}\begin{equation}
	\lambda_j^\mathrm{loc} (t) = \frac{ \sum_{m=1}^M \beta_m^{[R_j]} e^{- t / \tau_m} }{ \sum_{m=1}^M \beta_m^{[T]} e^{- t / \tau_m}}.
\end{equation}\end{linenomath*}
The results of such computation are plotted in Main Text Figure 2.

\subsection{Spatial distribution of feedback contributions}

\begin{figure}[p]
\centering
\includegraphics[width =\textwidth]{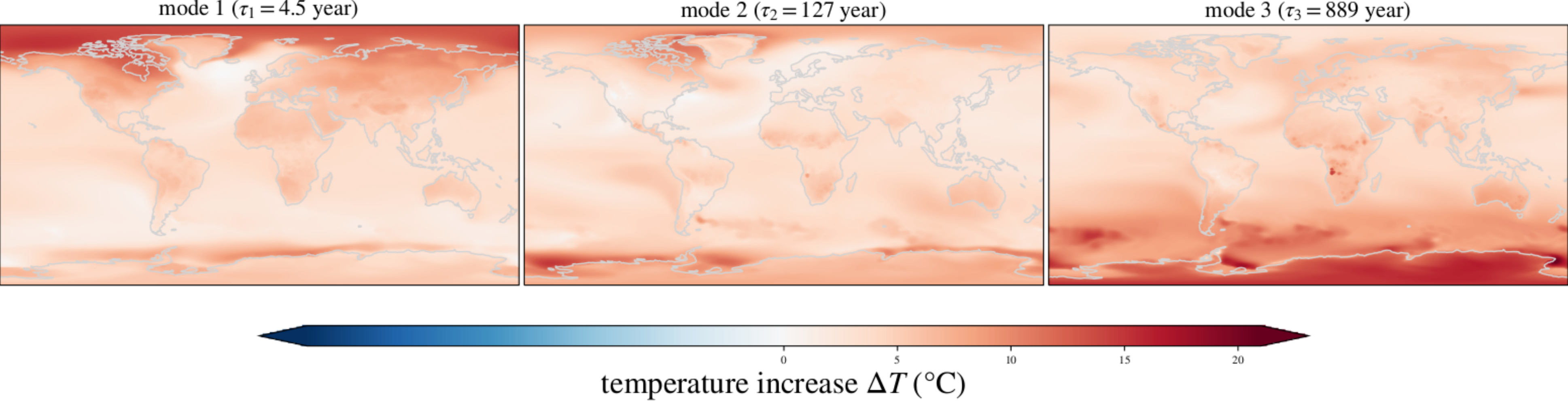}
\caption{Spatial distribution of temperature warming ($\Delta T$) modes.}
\label{fig:CESM2_spatial_DT}
\end{figure}

\begin{figure}
\centering
\includegraphics[width = \textwidth]{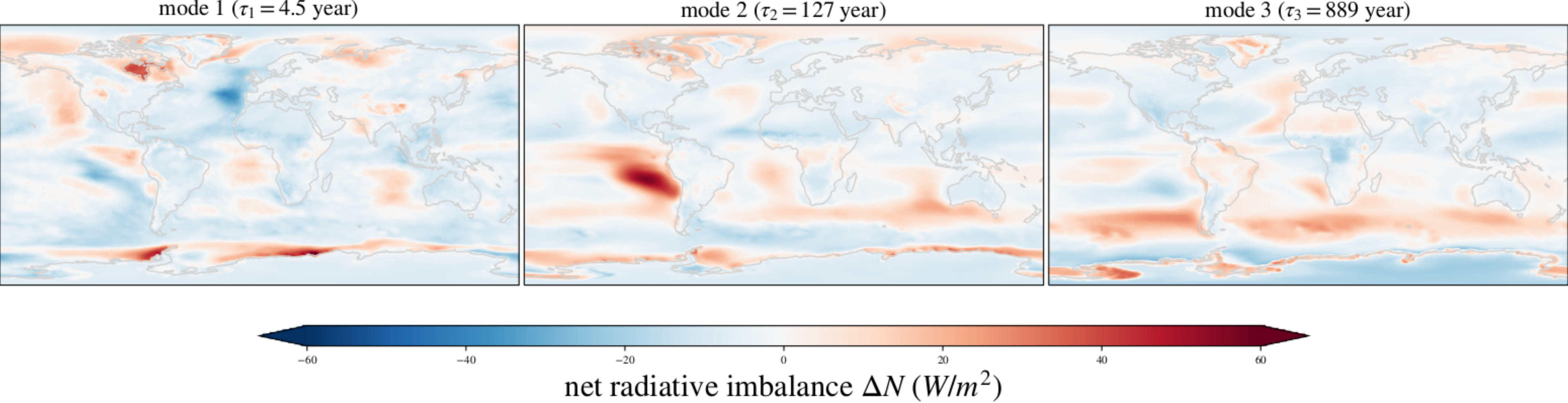}
\caption{Spatial distribution of net top-of-atmosphere radiative imbalance ($\Delta N$) modes.}
\label{fig:CESM2_spatial_N}
\end{figure}

\begin{figure}
\centering
\includegraphics[width = \textwidth]{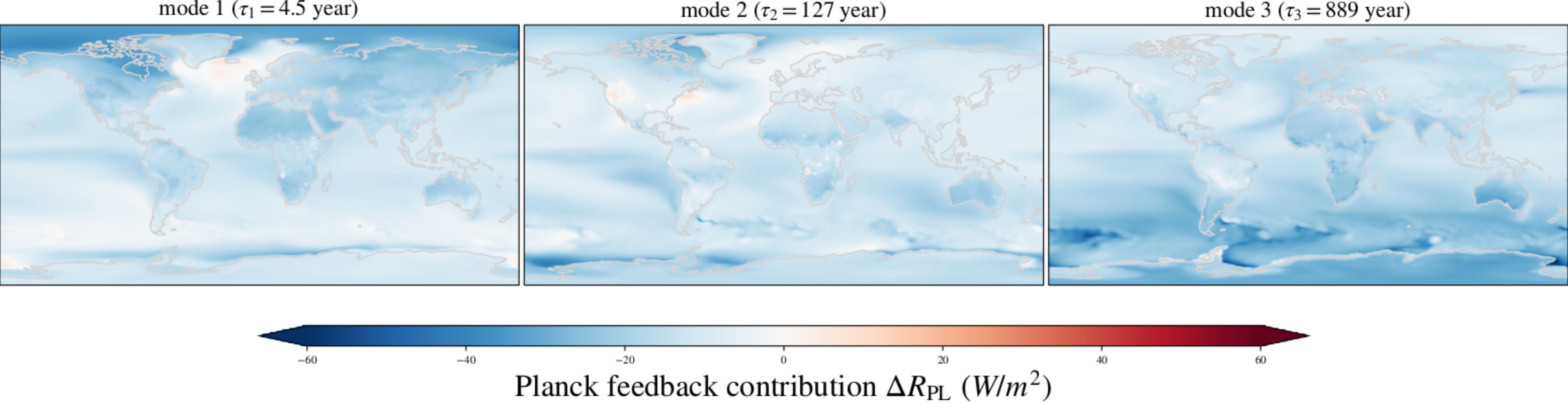}
\caption{Spatial distribution of Planck feedback contribution ($\Delta R_\textrm{PL}$) modes.}
\label{fig:CESM2_spatial_PL}
\end{figure}

\begin{figure}
\centering
\includegraphics[width =\textwidth]{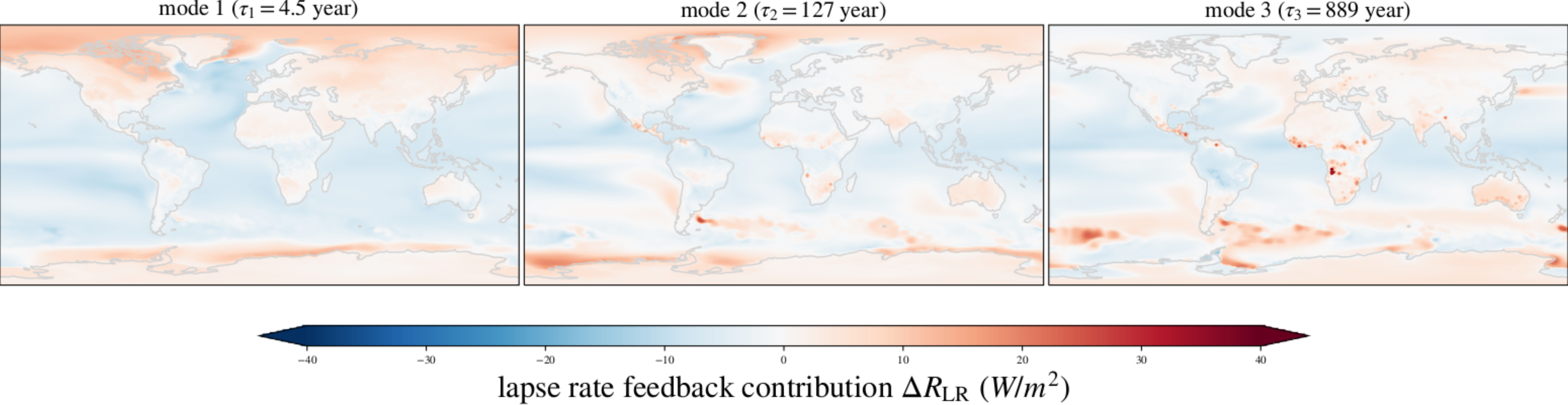}
\caption{Spatial distribution of Lapse Rate feedback contribution ($\Delta R_\textrm{LR}$) modes.}
\label{fig:CESM2_spatial_LR}
\end{figure}

\begin{figure}
\centering
\includegraphics[width = \textwidth]{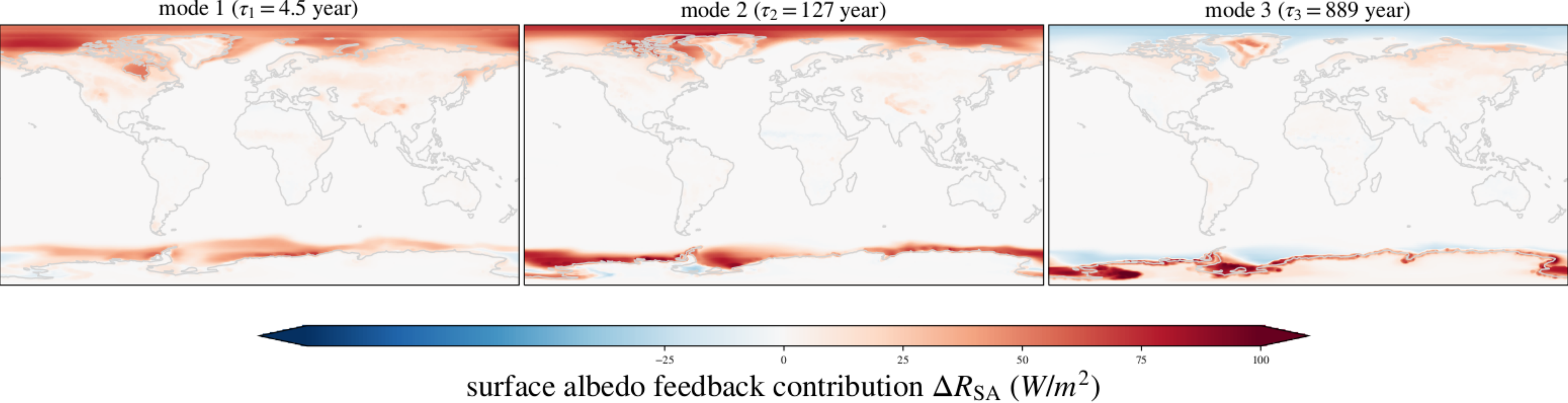}
\caption{Spatial distribution of Surface Albedo feedback contribution ($\Delta R_\textrm{SA}$) modes.}
\label{fig:CESM2_spatial_SA}
\end{figure}

\begin{figure}
\centering
\includegraphics[width = \textwidth]{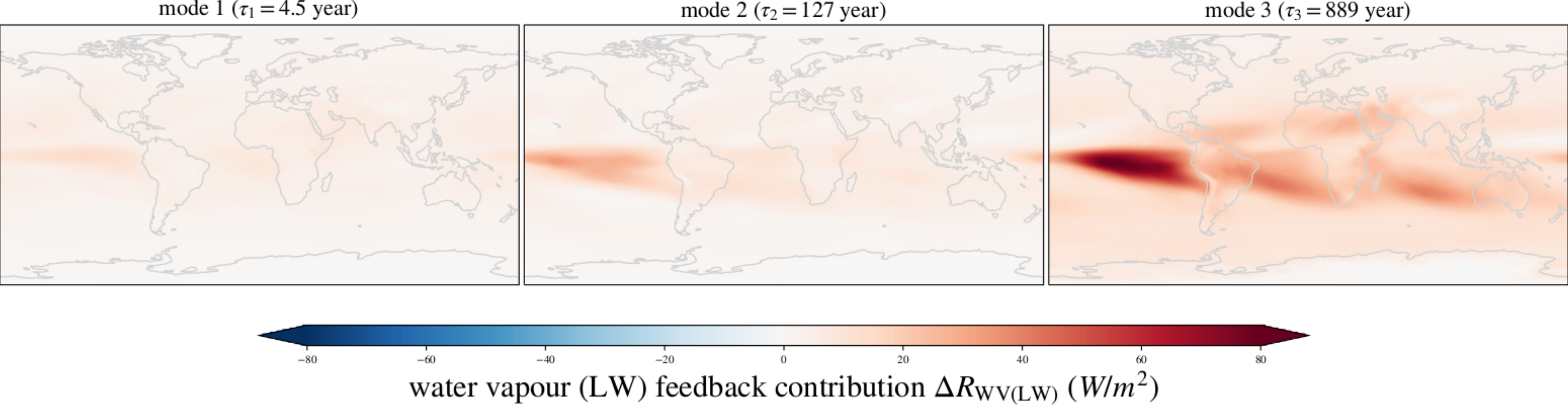}
\caption{Spatial distribution of Long Wave Water Vapour feedback contribution ($\Delta R_\textrm{WV(LW)}$) modes.}
\label{fig:CESM2_spatial_WVLW}
\end{figure}

\begin{figure}
\centering
\includegraphics[width = \textwidth]{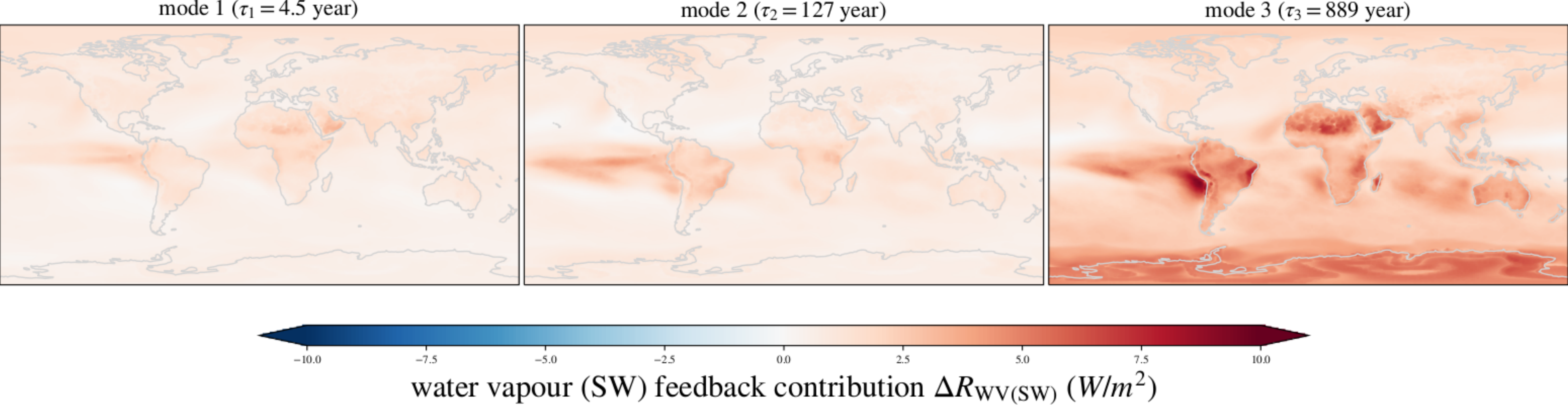}
\caption{Spatial distribution of Short Wave Water Vapour feedback contribution ($\Delta R_\textrm{WV(SW)}$) modes.}
\label{fig:CESM2_spatial_WVSW}
\end{figure}

\begin{figure}
\centering
\includegraphics[width = \textwidth]{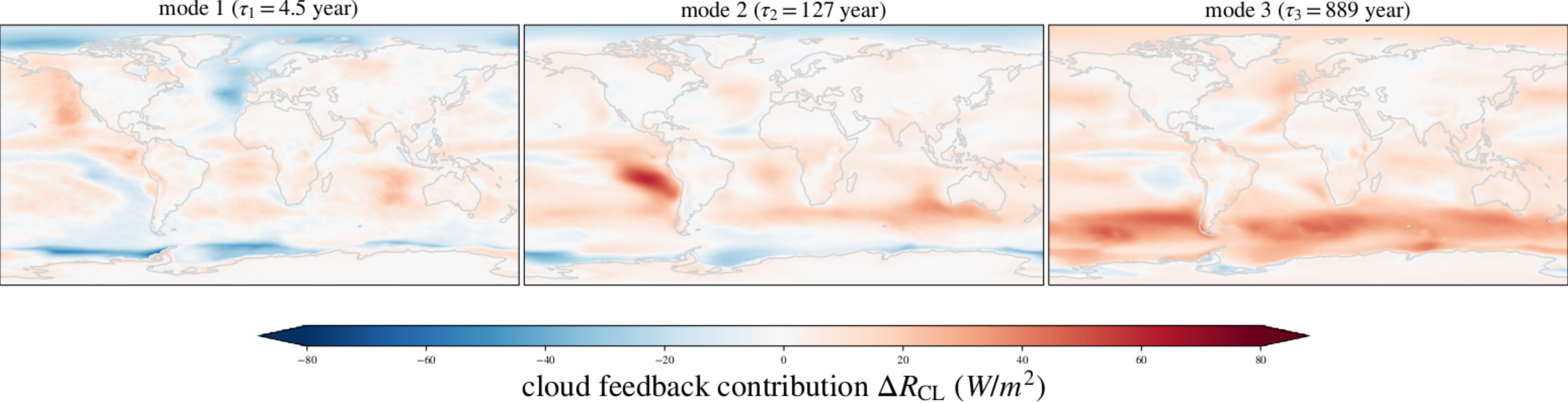}
\caption{Spatial distribution of Cloud feedback contribution ($\Delta R_\textrm{CL}$) modes.}
\label{fig:CESM2_spatial_CLOUD}
\end{figure}

To study the spatial distribution of the feedback contributions per mode, we consider an extension of equations \eqref{eq:CESM2_globalTemperature}-\eqref{eq:CESM2_globalImbalance}, in which both the time series and the mode contributions $\beta_m^{[O]}$ are considered per spatial location (cf. \cite{proistosescu2017slow}, where similar procedure has been applied to obtain spatial distribution of temperature warming modes). That is,
\begin{linenomath*}
\begin{align}
	\Delta T ({\bf x},t) = \sum_{m=1}^M \beta_m^{[T]}({\bf x})\ \mathcal{M}^{g_\textrm{abr}}_m(t) & = \sum_{m=1}^M \tilde{\beta}_m^{[T]}({\bf x})\ \left( 1 - e^{-t / \tau_m}\right); \\
	\Delta R_j ({\bf x}, t) = \sum_{m=1}^M \beta_m^{[j]}({\bf x})\ \mathcal{M}^{g_\textrm{abr}}_m(t) & = \sum_{m=1}^M \tilde{\beta}_m^{[j]} ({\bf x})\ \left( 1 - e^{-t / \tau_m} \right); \\
	\Delta N ({\bf x},t) = F({\bf x}) + \sum_{m=1}^M \beta_m^{[j]}({\bf x})\ \mathcal{M}^{g_\textrm{abr}}_m(t) & = F({\bf x}) + \sum_{m=1}^M \tilde{\beta}_m^{[j]} ({\bf x})\ \left( 1 - e^{-t / \tau_m} \right).
\end{align}
\end{linenomath*}
The values for the time scales $\tau_m$ as determined from the fits of the global variables can be used here. This makes finding the spatial eigenmodes possible via (multivariate) linear regression. We have performed this procedure on 2D fields for all feedback contributions $\Delta R_j({\bf x},t)$, surface temperature warming $\Delta T({\bf x},t)$ and net top-of-atmsophere radiative imbalance $\Delta N({\bf x},t)$. (3D fields have been averaged over height/pressure coordinate). The resulting spatial modes are given in Figures~\ref{fig:CESM2_spatial_DT}-\ref{fig:CESM2_spatial_CLOUD}. In addition, estimates for equilibrium spatial distributions for these observables are shown in Main Text Figure 2.


\FloatBarrier
\newpage

\section{Projections for the 1pctCO2 experiment}

Expression~\eqref{eq:observable_evolution} holds for any (small enough) forcing $g$. In particular, only $\mathcal{M}_m^g$ changes when the forcing $g$ changes, while $\beta_m^{[O]}$ does not change. Hence, it is possible to determine $\beta_m^{[O]}$ from one experiment (i.e. for some forcing $g$) and use that for projections in case of other forcings. This is a form of linear response theory~\citep{ruelle2009review, lucarini2011statistical, ragone2016new} that already has been used succesfully in projections of for example ensemble-average warming in global climate models~\citep{lembo2020beyond, aengenheyster2018point}. Here, we do not have such ensemble, but nevertheless apply this principle to illustrate its potency in projections of feedback contributions. Specifically, we use the previously found fits for the CESM2 abrupt-4xCO2 experiment to construct projections for its accompanying 1pctCO2 experiment -- an experiment in which the atmospheric CO$_2$ concentrations are increased by $1\%$ every year. As that experiment has been performed, we can compare the projections to the actual model output. To do so, we have computed feedback contributions $\Delta R_j$ and other observables using the same techniques as before with the abrupt-4xCO2 experiment.

To relate the response in the abrupt-4xCO2 experiment to the response in the 1pctCO2 experiment, $\mathcal{M}_m^g$ need to be computed for both experiments. For this, we denote the forcing in the abrupt4xCO2 setting by $g_\mathrm{abr}(t)$ and in the 1pctCO2 setting by $g_\mathrm{grad}(t)$ -- where `abr' stands for `abrupt' and `grad' for `gradual'.

The radiative forcing due to changes in atmospheric CO$_2$ content is related logarithmically~\cite{}. Hence, experienced (change in) forcing $g$ is given by
\begin{linenomath*}\begin{equation}
	g(t) = \mathcal{A} \log\left[ c(t) \right] - \mathcal{A} \log\left[ c(0) \right] = \mathcal{A} \log\left[ \frac{c(t)}{c(0)} \right],\label{eq:change_in_forcing}
\end{equation}\end{linenomath*}
where $\mathcal{A}$ is some constant (according to~\cite{Myhre2014}, $\mathcal{A} = 5.35\ W m^{-2}$ for the effect of CO$_2$ alone, but higher values are also used sometimes to mimic enslaved increase in other greenhouse gases~\citep{aengenheyster2018point}; in any case, the precise value is completely irrelevant for the computations here as it appears as multiplicative constant for all forcings) and where $c(t)$ denotes the concentration of CO$_2$ in the atmosphere over time. We note that the formalism presented here also would hold if $g$ was just the atmospheric CO$_2$ concentration, but then the `smallness of forcing' requirement is much stricter, making projections useless in practice (this can be seen `intuitively' by making a Taylor approximation of \eqref{eq:change_in_forcing} for small changes in atmospheric CO$_2$).

Now, $\mathcal{M}_m^{g}$ can be computed straightforwardly. Since $c_\mathrm{abr}(t) = 4 c(0)$, $g_\mathrm{abr}(t) = \mathcal{A} \log(4)$ and hence
\begin{linenomath*}\begin{equation}
	\mathcal{M}_m^{g_\mathrm{abr}} (t) = \mathcal{A} \log(4) \tau_m \left( 1 - e^{- t / \tau_m}\right),
\end{equation}\end{linenomath*}
as we found in~\eqref{eq:Mmg_for_abrupt} -- where $\Delta \mu = \mathcal{A} \log(4)$. In the 1pctCO2 scenario, $c_\mathrm{grad}(t) = c(0) \cdot (1.01)^t$. Therefore, $g_\mathrm{grad}(t) = \mathcal{A} \log(1.01) t$ and hence
\begin{linenomath*}
\begin{align}
	\mathcal{M}_m^{g_\mathrm{grad}}(t)
&= \mathcal{A} \log(1.01) \int_0^t e^{-s/\tau_m} (t-s) ds \nonumber\\
&= \mathcal{A} \log(1.01) \tau_m \left[ t + \tau_m \left( e^{-t / \tau_m} - 1 \right) \right].
\end{align}
\end{linenomath*}

\begin{figure}[p]
\centering
\includegraphics[width = \textwidth]{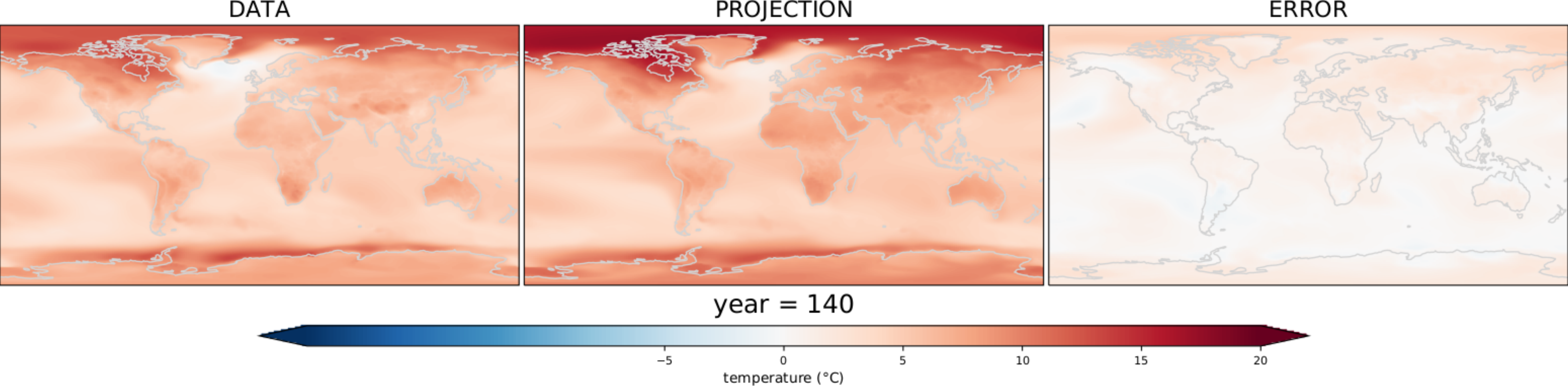}
\caption[1pctCO2 spatial projection of warming $\Delta T({\bf x})$ at year $140$]{Spatial distribution of warming $\Delta T({\bf x})$ at year $140$ of the 1pctCO2 experiment. Left panel shows the actual distribution computed from model output and the middle panel shows the projection made in this study. The right panel gives the difference between the projection and the actual values (i.e. red indicates an overestimation and blue an underestimation in the projection). A movie showing the evolution over the years is also available (SI-videos).}
\label{fig:projection_temperature}
\end{figure}

\begin{figure}
\centering
\includegraphics[width = \textwidth]{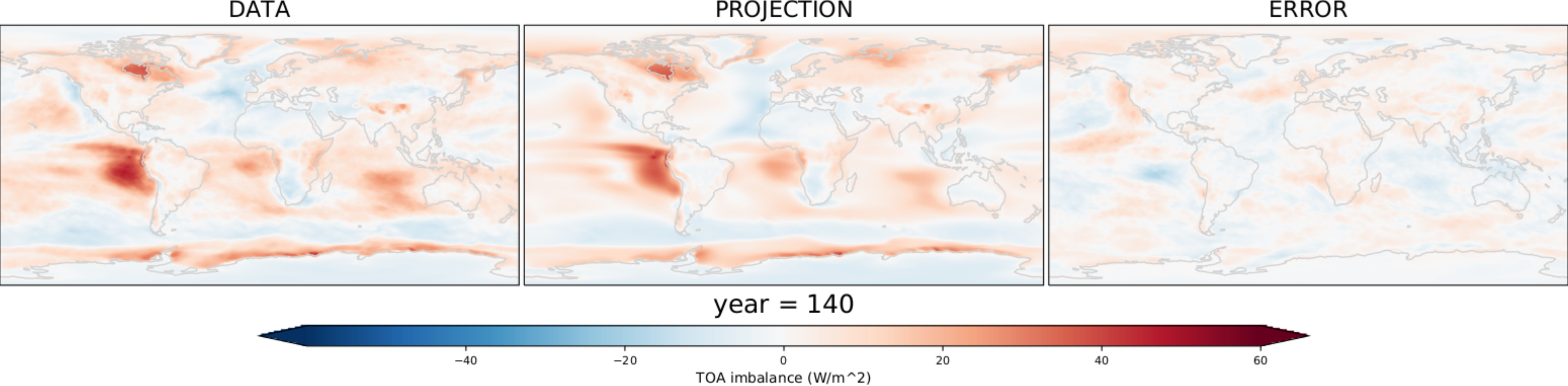}
\caption[1pctCO2 spatial projection of radiative imbalance $\Delta N({\bf x})$ at year $140$]{Spatial distribution of top-of-atmospher imbalance $\Delta N({\bf x})$ at year $140$ of the 1pctCO2 experiment. Left panel shows the actual distribution computed from model output and the middle panel shows the projection made in this study. The right panel gives the difference between the projection and the actual values (i.e. red indicates an overestimation and blue an underestimation in the projection). A movie showing the evolution over the years is also available (SI-videos).}
\label{fig:projection_imbalance}
\end{figure}

\begin{figure}
\centering
\includegraphics[width = \textwidth]{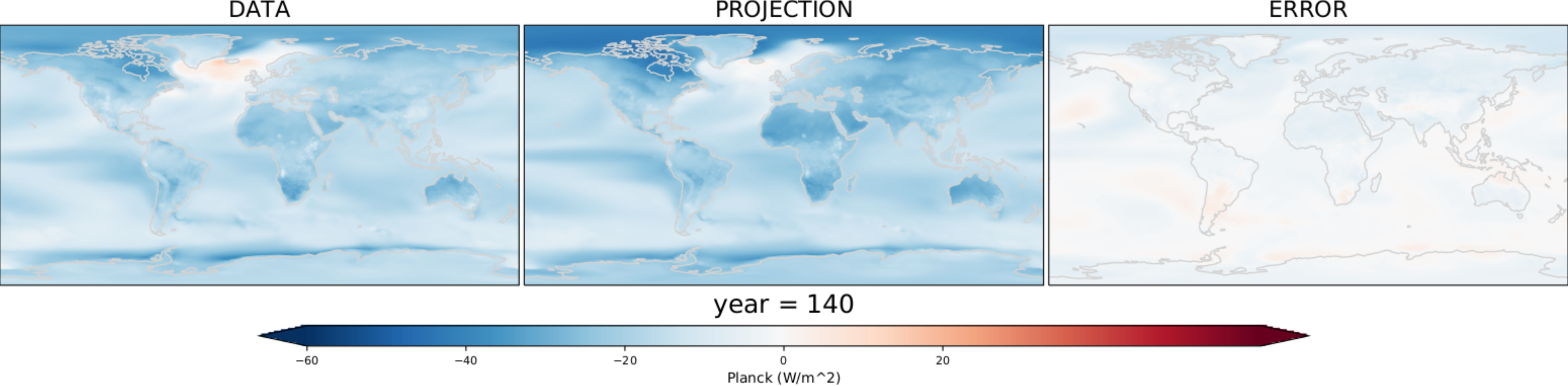}
\caption[1pctCO2 spatial projection of Planck feedback contribution $\Delta R_\mathrm{Planck}({\bf x})$ at year $140$]{Spatial distribution of Planck feedback contribution $\Delta R_\mathrm{Planck}({\bf x})$ at year $140$ of the 1pctCO2 experiment. Left panel shows the actual distribution computed from model output and the middle panel shows the projection made in this study. The right panel gives the difference between the projection and the actual values (i.e. red indicates an overestimation and blue an underestimation in the projection). A movie showing the evolution over the years is also available (SI-videos).}
\label{fig:projection_Planck}
\end{figure}

\begin{figure}
\centering
\includegraphics[width = \textwidth]{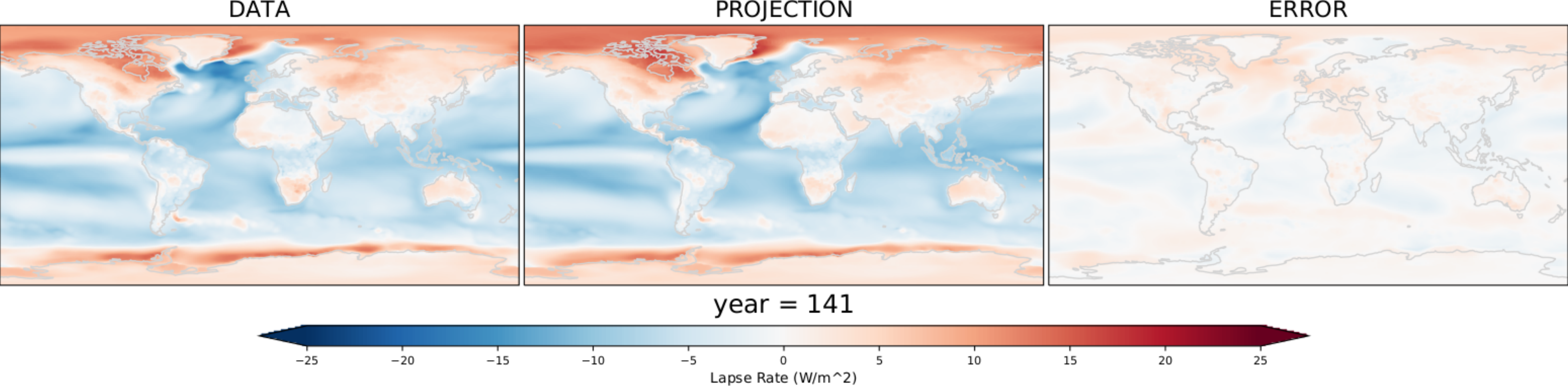}
\caption[1pctCO2 spatial projection of lapse rate feedback contribution $\Delta R_\mathrm{LR}({\bf x})$ at year $140$]{Spatial distribution of lapse rate feedback contribution $\Delta R_\mathrm{LR}({\bf x})$ at year $140$ of the 1pctCO2 experiment. Left panel shows the actual distribution computed from model output and the middle panel shows the projection made in this study. The right panel gives the difference between the projection and the actual values (i.e. red indicates an overestimation and blue an underestimation in the projection). A movie showing the evolution over the years is also available (SI-videos).}
\label{fig:projection_lapse_rate}
\end{figure}

\begin{figure}
\centering
\includegraphics[width = \textwidth]{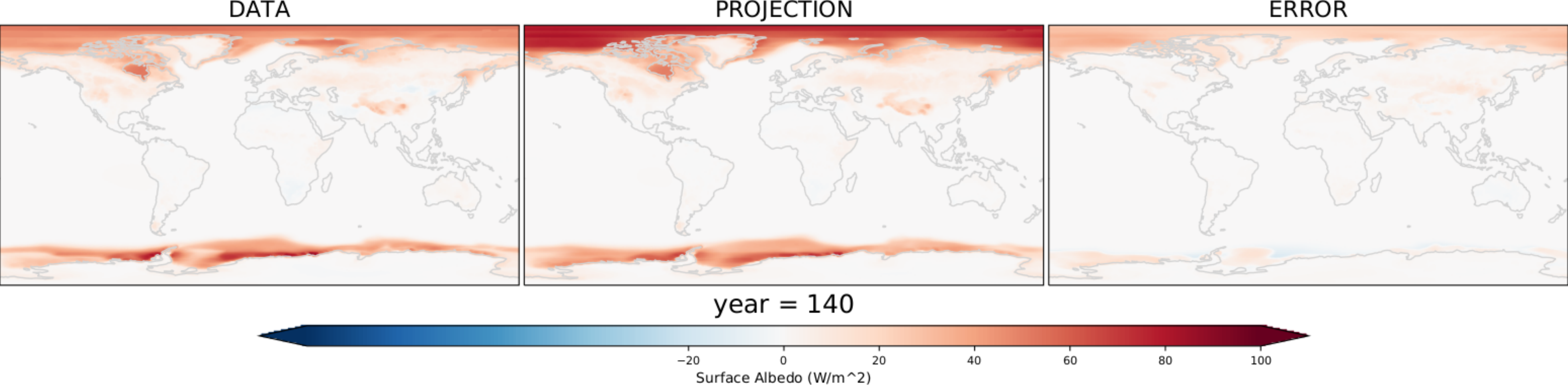}
\caption[1pctCO2 spatial projection of surface albedo feedback contribution $\Delta R_\mathrm{SA}({\bf x})$ at year $140$]{Spatial distribution of surface albedo feedback contribution $\Delta R_\mathrm{SA}({\bf x})$ at year $140$ of the 1pctCO2 experiment. Left panel shows the actual distribution computed from model output and the middle panel shows the projection made in this study. The right panel gives the difference between the projection and the actual values (i.e. red indicates an overestimation and blue an underestimation in the projection). A movie showing the evolution over the years is also available (SI-videos).}
\label{fig:projection_albedo}
\end{figure}

\begin{figure}
\centering
\includegraphics[width = \textwidth]{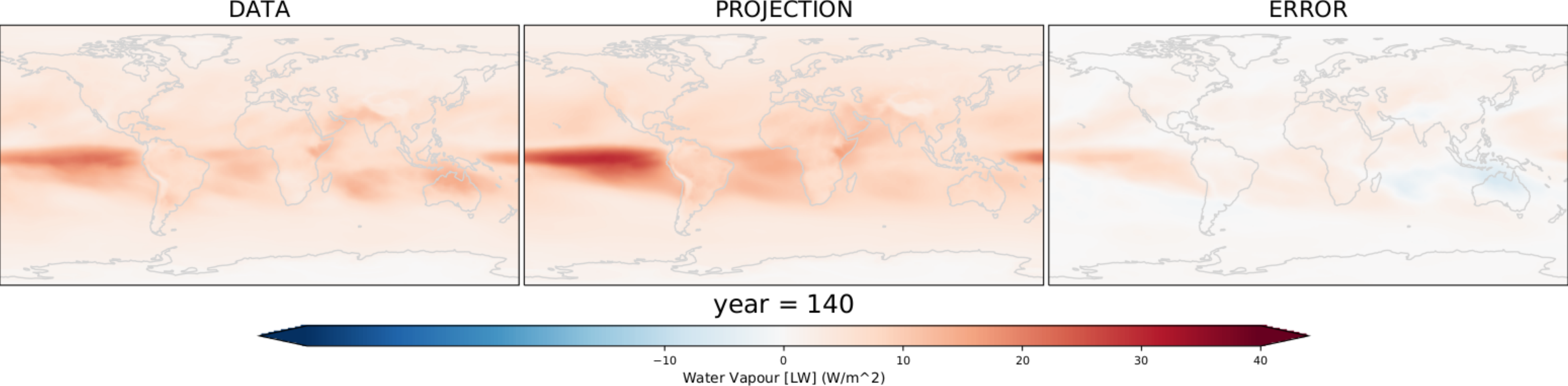}
\caption[1pctCO2 spatial projection of LW water vapour feedback contribution $\Delta R_\mathrm{WV-LW}({\bf x})$ at year $140$]{Spatial distribution of long-wave water vapour feedback contribution $\Delta R_\mathrm{WV-LW}({\bf x})$ at year $140$ of the 1pctCO2 experiment. Left panel shows the actual distribution computed from model output and the middle panel shows the projection made in this study. The right panel gives the difference between the projection and the actual values (i.e. red indicates an overestimation and blue an underestimation in the projection). A movie showing the evolution over the years is also available (SI-videos).}
\label{fig:projection_water_vapour_LW}
\end{figure}

\begin{figure}
\centering
\includegraphics[width = \textwidth]{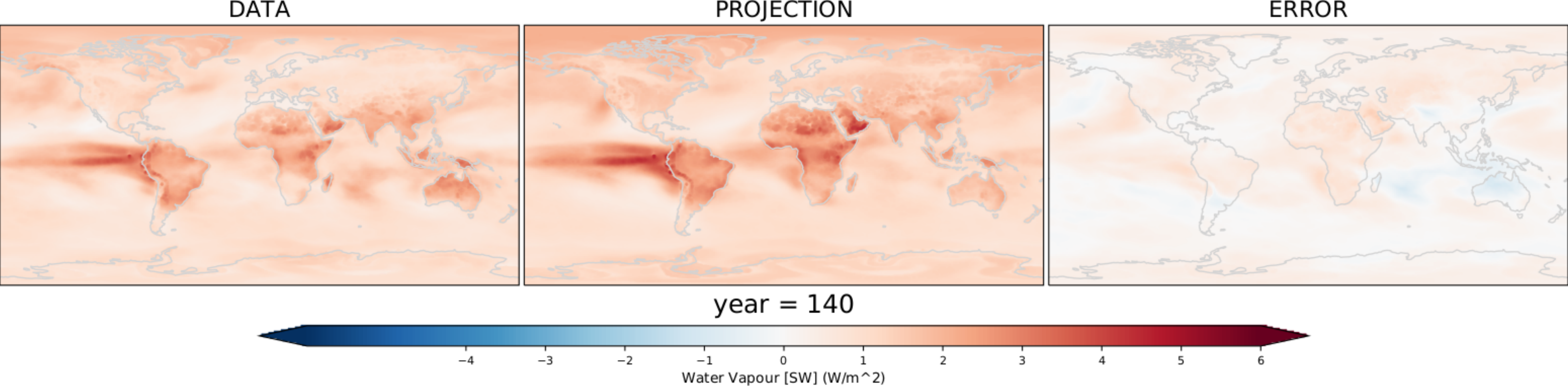}
\caption[1pctCO2 spatial projection of SW water vapour feedback contribution $\Delta R_\mathrm{WV-SW}({\bf x})$ at year $140$]{Spatial distribution of short-wave water vapour feedback contribution $\Delta R_\mathrm{WV-SW}({\bf x})$ at year $140$ of the 1pctCO2 experiment. Left panel shows the actual distribution computed from model output and the middle panel shows the projection made in this study. The right panel gives the difference between the projection and the actual values (i.e. red indicates an overestimation and blue an underestimation in the projection). A movie showing the evolution over the years is also available (SI-videos).}
\label{fig:projection_water_vapour_SW}
\end{figure}

\begin{figure}
\centering
\includegraphics[width = \textwidth]{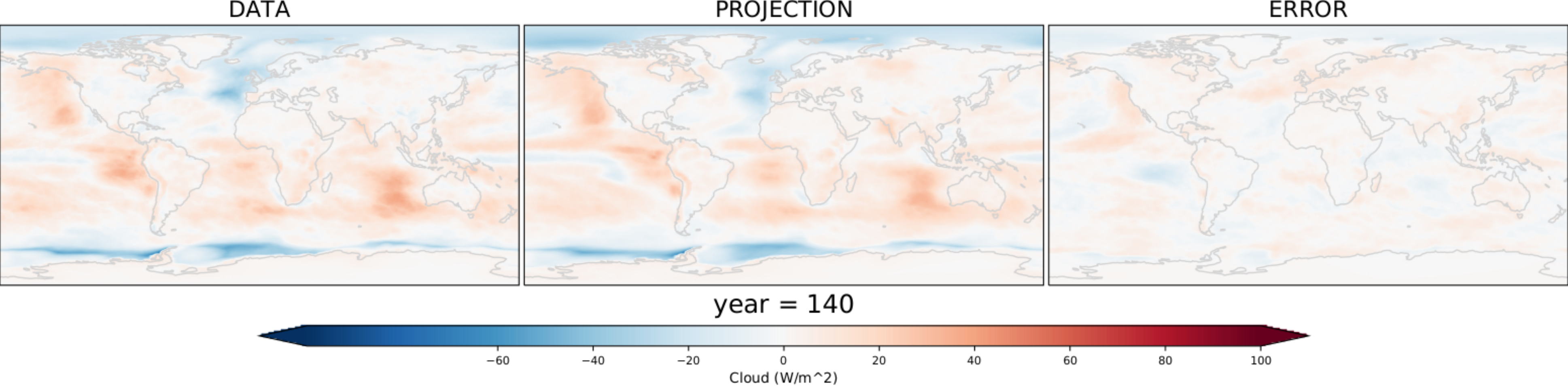}
\caption[1pctCO2 spatial projection of cloud feedback contribution $\Delta R_\mathrm{Cloud}({\bf x})$ at year $140$]{Spatial distribution of cloud feedback contribution $\Delta R_\mathrm{Cloud}({\bf x})$ at year $140$ of the 1pctCO2 experiment. Left panel shows the actual distribution computed from model output and the middle panel shows the projection made in this study. The right panel gives the difference between the projection and the actual values (i.e. red indicates an overestimation and blue an underestimation in the projection). A movie showing the evolution over the years is also available (SI-videos).}
\label{fig:projection_clouds}
\end{figure}

With these expressions, projections for the feedback contributions $\Delta R_j$ (as well as for warming $\Delta T$ and radiative imbalance $\Delta N$) in the 1pctCO2 experiment can be made. The projections for the globally averaged observables are given in Main Text Figure 3 along with the computed actual values. Spatial projections are also possible; in Figures~\ref{fig:projection_temperature}--\ref{fig:projection_clouds} the spatial projections at $t = 140y$ (at that moment, the amount of CO$_2$ has been quadruppled compared to $t = 0$) are given along with the actual values and the errors (the spatial distribution for surface albedo feedback also appears as Main Text Figure 4). Movies of the spatial projections over the years are also available.


As before, the instantaneous feedback strengths $\lambda_j^\mathrm{inst}(t)$ can be computed. For the 1pctCO2 experiment these are given by
\begin{linenomath*}\begin{equation}
	\lambda_j^\mathrm{inst,grad}(t) = \frac{ \sum_{m=1}^M \beta_m^{[R_j]} \tau_m \left( 1 - e^{ - t / \tau_m } \right) }{ \sum_{m=1}^M \beta_m^{[T]} \tau_m \left( 1 - e^{ - t / \tau_m } \right)} = \frac{\Delta R_j^\mathrm{abr}(t) }{ \Delta T^\mathrm{abr}(t)} = \lambda_j^\mathrm{abr}(t).
\end{equation}\end{linenomath*}
That is, the intantaneous feedback strength in the gradual setting corresponds to the fraction $\Delta R_j(t) / \Delta T(t)$ in the abrupt setting\footnote{This is not a coincidence and does not depend on the form of the Green function $G^{[O]}$, because $\frac{d}{dt} \left( G^{[O]} * g^{grad}\right)(t) = \left( G^{[O]} * \frac{d g^\mathrm{grad}}{dt}\right)(t) = K \left( G^{[O]} * g^\mathrm{abr}\right)(t)$, where $K = \log(4)/\log(1.01)$.} -- and hence to the classical feedback strength in that setting, as per~\eqref{eq:climate_feedback}. A plot of $\lambda_j^\mathrm{loc,grad}(t)$ is shown in Main Text Figure 2


\newpage
\section{Cloud feedback}
\label{sec:cloudFeedback}

Following \cite{soden2008quantifying}, the cloud feedbacks have been derived from differences between clearsky and fullsky feedback contributions of the other feedbacks. As the total radiative response is the sum over all feedback contributions, we have
\begin{linenomath*}
\begin{align}
	\Delta N(t)  = F(t) + \Delta R(t) & = F(t) + \sum_{j \in \mathcal{F}} \Delta R_j(t).
	\label{eq:cloudfeedback_fullsky}
\intertext{
Similarly, the clearsky radiative (im)balance constitutes of all feedbacks except for the cloud feedback, i.e.
}
	\Delta N^\mathrm{cs}(t) = F^\mathrm{cs}(t) + \Delta R^\mathrm{cs}(t) & = F^\mathrm{cs}(t) + \sum_{j \in \widetilde{\mathcal{F}} } \Delta R^\mathrm{cs}_j(t),
	\label{eq:cloudfeedback_clearsky}
\end{align}
\end{linenomath*}
where the superscript `$\mathrm{cs}$' refers to the clear sky values, $\widetilde{\mathcal{F}}$ is the set of all feedback mechanisms except for the cloud feedback -- that does play no role in the clear sky setting -- and $R^\mathrm{cs}_j(t)$ is the feedback contribution of the $j$-th feedback in the clear sky scenario, i.e.
\begin{linenomath*}\begin{equation}
	\Delta R_j^\mathrm{cs}(t) := \frac{\partial \Delta N^\mathrm{cs}}{\partial y_j} \Delta y_j(t).
\end{equation}\end{linenomath*}
Upon using the tautology $\Delta N(t) = \Delta N^\mathrm{cs}(t) + \left[\Delta N(t) - \Delta N^\mathrm{cs}(t) \right]$, \eqref{eq:cloudfeedback_fullsky} and \eqref{eq:cloudfeedback_clearsky} can be used to obtain
\begin{linenomath*}\begin{equation}
	F^\mathrm{cs}(t) + \sum_{j \in \widetilde{\mathcal{F}}} \Delta R_j^\mathrm{cs}(t) + \left[ \Delta N(t) - \Delta N^\mathrm{cs}(t)\right] = F(t) + \sum_{j \in \widetilde{\mathcal{F}}} \Delta R_j(t) + \Delta R_\mathrm{cloud}(t).
\end{equation}\end{linenomath*}
Rearranging these terms yields the cloud feedback contribution
\begin{linenomath*}\begin{equation}
	\Delta R_\mathrm{cloud}(t) = \left[ \Delta N(t) - \Delta N^\mathrm{cs}(t) \right] + \left[ F^\mathrm{cs}(t) - F(t) \right] + \sum_{j \in \widetilde{\mathcal{F}}} \left[ \Delta R_j^\mathrm{cs}(t) - \Delta R_j(t)\right].
\end{equation}\end{linenomath*}
Because the radiative forcings $F(t)$ and $F^\mathrm{cs}(t)$ cannot be obtained from model output like $\Delta N(t)$, $\Delta N^\mathrm{cs}(t)$, $\Delta R_j(t)$ and $\Delta R_j^\mathrm{cs}(t)$, their values need to be derived from fits instead. Hence, in the fits we use the following expression:
\begin{linenomath*}\begin{equation}
	\left[ \Delta N(t) - \Delta N^\mathrm{cs}(t) \right] + \sum_{j \in \overline{\mathcal{F}}} \left[ \Delta R_j^\mathrm{cs}(t) - \Delta R_j(t) \right] = \left[ F(t) - F^\mathrm{cs}(t) \right] + \Delta R_\mathrm{Cloud}(t) = \left[ F(t) - F^\mathrm{cs}(t) \right] + \sum_{m=1}^M \beta_m^{[\mathrm{cloud}]} \mathcal{M}_m(t).
\end{equation}\end{linenomath*}


\bibliographystyle{abbrvnat}
\bibliography{../sources}